\documentclass{iopart}
\usepackage{graphicx}  
\usepackage{latexsym}
\usepackage{amssymb}

\jl{6}        
\eqnobysec    

\def\beq{\begin{equation}}
\def\eeq{\end{equation}}
\def\rmd{{\rm d}} 

\begin{document}

\title[Tidal indicators in the spacetime of a rotating deformed mass]
{Tidal indicators in the spacetime of a rotating deformed mass}

\author{
Donato Bini$^* {}^\S{}^\dag$, 
Kuantay Boshkayev${}^\S{}^\ddag{}^\P$ and 
Andrea Geralico${}^\S{}^\ddag$ 
}

\address{${}^*$\
Istituto per le Applicazioni del Calcolo ``M. Picone,'' CNR, 
I--00185 Rome, Italy
}

\address{
  ${}^\S$\
  ICRA,
  University of Rome ``La Sapienza,'' I--00185 Rome, Italy
}

\address{
  ${}^\dag$\
  INFN, Sezione di Firenze, I--00185
  Sesto Fiorentino (FI), Italy
}

\address{
  $^\ddag$
  Physics Department,
  University of Rome ``La Sapienza,'' I--00185 Rome, Italy
}

\address{${}^\P$\
ICRANet, Piazzale della Repubblica 10, I-65122 Pescara, Italy
}

\ead{binid@icra.it}

\begin{abstract}
Tidal indicators are commonly associated with the electric and magnetic parts of the Riemann tensor (and its covariant derivatives) with respect to a given family of observers in a given spacetime. 
Recently, observer-dependent tidal effects have been extensively investigated with respect to a variety of special observers in the equatorial plane of the Kerr spacetime.
This analysis is extended here by considering a more general background solution to include the case of matter which is also endowed with an arbitrary mass quadrupole moment.
Relation with curvature invariants and Bel-Robinson tensor, i.e., observer-dependent super-energy density and super-Poynting vector, are investigated too.
\end{abstract}

\pacno{04.20.Cv}

\section{Introduction}

Relativistic tidal problems have been extensively studied in the literature in a variety of contexts.
Tidal effects are responsible for deformations or even disruption of astrophysical objects (like ordinary stars but also compact objects like neutron stars) placed in strong gravitational fields.
For instance, they play a central role in the merging of compact object binaries, which can be accurately modeled only by numerical simulations in full general relativity, solving the coupled Einstein-hydrodynamics equations needed to evolve relativistic, self-gravitating fluids \cite{fullgr1,fullgr2,fullgr3}.
Such tidal disruption events are expected to happen very frequently in the Universe, leading then to a possible detection of the associated emission of gravitational waves in the near future by ground-based detectors \cite{abadieetal}.
To this end, different analytical and semi-analytical approaches have been developed to properly describe at least part of the coalescence process and to study the associated gravitational wave signals \cite{fla-hin,hin1,damour1}.
These approaches usually require either Post-Newtonian techniques or first order perturbation theory.
In fact, in this limit the motion of each individual compact object in the binary system can be treated as the motion of an extended body in a given gravitational field due to its companion under the assumption that it causes only a small perturbation on the background \cite{mash75}.
Finally, one can also be interested in studying the tidal disruption limit of ordinary stars and compact objects in the field of a black hole (see, e.g., Ref. \cite{mino} and references therein).
Tidal problems of this kind can be treated within the so-called tidal approximation, i.e., by assuming that the mass of the star is much smaller than the black hole mass and that the stellar radius is smaller than the orbital radius, so that backreaction effects on the background field can be neglected.
Therefore, the star is usually described as a self-gravitating Newtonian fluid and its center of mass is assumed to move around the black hole along a timelike geodesic path.
The tidal field due to the black hole is then computed from the Riemann tensor in terms of the geodesic deviation equation.

The role of the observer in relativistic tidal problems has never received enough attention in the literature.
Nevertheless, it is crucial in interpreting the results.
In fact, if tidal forces are due to curvature, the latter is experienced by observers through the electric and magnetic parts of the Riemann tensor, which is the only true $4$-dimensional invariant quantity.
In contrast, its electric and magnetic parts depend by definition on the choice of the observers who perform the measurement. 
In a recent paper \cite{kerr_tidal} we have addressed such an issue by providing all necessary tools to relate the measurement of tidal effects by different families of observers. 
We have considered two \lq\lq tidal indicators" defined as the trace of the square of the electric and magnetic parts of the Riemann tensor, respectively.
They are both curvature and observer dependent and we have investigated their properties by considering a number of special observer families in the equatorial plane of the Kerr spacetime.
As an interesting feature we have shown that the electric-type indicator cannot be made as vanishing with respect to any such observers, whereas the family of Carter's observers is the only one which measures zero tidal magnetic indicator. 
We have argued that the explanation for this effect relies on the absence of a quadrupole moment independent on the rotational parameter for the Kerr solution. To answer this question, as well as to extend the previous analysis to a more general context, we consider here a solution of the vacuum Einstein field equations due to Quevedo and Mashhoon \cite{quev89,quevmas91}, which generalizes the Kerr spacetime to include the case of matter with arbitrary mass quadrupole moment and is specified by three parameters, the mass $M$, the angular momentum per unit mass $a$ and the quadrupole parameter $q$.
It is its genuine quadrupole moment content which makes this solution of particular interest here. 
We will thus investigate how the shape deformation of the rotating source affects the properties of tidal indicators with respect to special family of observers, including static observers, ZAMOs (i.e., zero angular momentum observers) and geodesic observers.

Hereafter latin indices run from $1$ to $3$ whereas greek indices run from $0$ to $3$ and geometrical units are assumed. The metric signature is chosen 
as $+2$.

\section{The gravitational field of a rotating deformed mass}

The exterior gravitational field of a rotating deformed mass can be described by the Quevedo-Mashhoon (hereafter QM) solution \cite{quev89,quevmas91}.
This is a stationary axisymmetric solution of the vacuum Einstein's equations belonging to the class of  
Weyl-Lewis-Papapetrou \cite{weyl17,lew32,pap66} and is characterized, in general, by the presence of a naked singularity.
Although the general solution is characterized by an infinite set of gravitoelectric and gravitomagnetic multipoles, we consider here the special solution discussed in Ref. \cite{quevmas85} that involves only three parameters: the mass $M$, the angular momentum per unit mass $a$ and the mass quadrupole parameter $q$ of the source.

The corresponding line element in prolate spheroidal coordinates ($t,x,y,\phi$) with $x \geq 1$, $-1 \leq y \leq 1$ is given by \cite{ES}
\begin{eqnarray}\fl\quad
\label{metgen}
\rmd s^2&=&-f(\rmd t-\omega \rmd\phi)^2\nonumber\\
\fl\quad
&&+\frac{\sigma^2}{f}\left\{e^{2\gamma}\left(x^2-y^2\right)\left(\frac{\rmd x^2}{x^2-1}+\frac{\rmd y^2}{1-y^2}\right)+(x^2-1)(1-y^2)\rmd\phi^2\right\}\ ,
\end{eqnarray}
where $f$, $\omega$ and $\gamma$ are functions of $x$ and $y$ only and $\sigma$ is a constant.
They have the form
\begin{eqnarray}
f&=&\frac{R}{L} e^{-2qP_2Q_2}\ , \qquad
\omega=-2a-2\sigma\frac{\mathfrak M}{R} e^{2qP_2Q_2}\ , \nonumber\\
e^{2\gamma}&=&\frac{1}{4}\left(1+\frac{M}{\sigma}\right)^2\frac{R}{x^2-1} e^{2\hat\gamma}\ ,
\end{eqnarray}
where
\begin{eqnarray}\fl\quad
\label{variousdefs}
R&=& a_+ a_- + b_+b_-\ , \qquad 
L = a_+^2 + b_+^2\ , \nonumber\\
\fl\quad
{\mathfrak M}&=&\alpha x(1-y^2)(e^{2q\delta_+}+e^{2q\delta_-}) a_+ +y(x^2-1)(1-\alpha^2e^{2q(\delta_++\delta_-)})b_+\ , \nonumber\\
\fl\quad
\hat\gamma&=&\frac12(1+q)^2 \ln\frac{x^2-1}{x^2-y^2} + 2q(1-P_2)Q_1 + q^2(1-P_2) \bigg[ (1+P_2)(Q_1^2-Q_2^2)\nonumber \\
\fl\quad
&&+\frac12(x^2-1)(2Q_2^2 - 3xQ_1Q_2 + 3 Q_0Q_2 - Q_2')\bigg] \ .
\end{eqnarray}
Here $P_l(y)$ and $Q_l(x)$ are Legendre polynomials of the first and second kind respectively. 
Furthermore
\begin{eqnarray}\fl\quad
\label{variousdefs2}
a_\pm &=& x( 1-\alpha^2e^{2q(\delta_++\delta_-)})\pm ( 1+\alpha^2e^{2q(\delta_++\delta_-)})\ , \nonumber\\
\fl\quad
b_\pm &=& \alpha y ( e^{2q\delta_+}+e^{2q\delta_-}) \mp \alpha ( e^{2q\delta_+}- e^{2q\delta_-}) \ , \nonumber\\
\fl\quad
\delta_\pm &=& \frac12\ln\frac{(x\pm y)^2}{x^2-1} +\frac32 (1-y^2\mp xy)+\frac{3}{4}[x(1-y^2) \mp y (x^2-1)]\ln\frac{x-1}{x+1}\ ,
\end{eqnarray}
the quantity $\alpha$ being a constant
\beq
\label{metquev}
\alpha=\frac{\sigma-M}{a}\ , \qquad \sigma = \sqrt{M^2-a^2}\ .
\eeq
We limit our analysis here to the case $\sigma>0$, i.e. $M>a$. 
In the case $\sigma=0$ the solution reduces to the extreme Kerr spacetime irrespective of the value of $q$ \cite{quevmas91}.

The  Geroch-Hansen \cite{ger,hans} moments are given by
\begin{equation} 
M_{2k+1} = J_{2k}=0 \ ,  \quad k = 0,1,2,... 
\end{equation} 
\begin{equation} 
M_0 = M \ , \quad M_2 = - Ma^2 + \frac{2}{15}qM^3 \left(1-\frac{a^2}{M^2}\right)^{3/2}  \ , ... 
\label{elemm}
\end{equation} 
\begin{equation} 
J_1= Ma \ , \quad J_3 = -Ma^3 +  \frac{4}{15}qM^3 a \left(1-\frac{a^2}{M^2}\right)^{3/2}  \ , ....
\label{magmm}
\end{equation} 
The vanishing of the odd gravitoelectric ($M_n$) and even gravitomagnetic ($J_n)$ multipole moments is a consequence of the reflection symmetry of the solution about the hyperplane $y=0$, which we will refer to as \lq\lq symmetry'' (or equivalently \lq\lq equatorial'') plane hereafter. 
From the above expressions we see that $M$ is the total mass of the body, $a$ represents the 
specific angular momentum, and $q$  is related to the deviation from spherical symmetry. All higher multipole moments can be shown to depend only on the parameters $M$, $a$, and $q$ \cite{quevmas85}. 

Some geometric and physical properties of the QM solution have been analyzed in Ref. \cite{bglq}.  
It turns out that the whole geometric structure of the QM spacetime is drastically changed in comparison with Kerr spacetime, leading to a number of previously unexplored physical effects strongly modifying the features of particle motion, especially near the gravitational source. 
In fact, the QM solution is characterized by a naked singularity at $x=1$, whose existence critically depends on the value of the quadrupole parameter $q$. In the case $q=0$ (Kerr solution) $x=1$ represents instead an event horizon.

\subsection{Limiting cases}
\label{sec:spe}

The QM solution reduces to the Kerr spacetime in the limiting case $q\rightarrow 0$, to the Erez-Rosen spacetime when $a\rightarrow 0$ and to the Schwarzschild solution when both parameters vanish.
Furthermore, it has been shown in Ref. \cite{bglq} that the general form of the QM solution is equivalent, up to a coordinate transformation, to the exterior vacuum Hartle-Thorne solution once linearized to first order in the quadrupole parameter and to second order in the rotation parameter. 
The limiting cases contained in the general solution thus suggest that it can be used to describe the exterior asymptotically flat gravitational field of a rotating body with arbitrary quadrupole moment.

\subsubsection{Kerr solution}

For vanishing quadrupole parameter we recover the Kerr solution, with functions \cite{ES}
\begin{eqnarray}
f_K&=&\frac{c^2x^2+d^2y^2-1}{(cx+1)^2+d^2y^2}\ , \qquad 
\omega_K=2a\frac{(cx+1)(1-y^2)}{c^2x^2+d^2y^2-1}\ , \nonumber\\
\gamma_K&=&\frac12\ln\left(\frac{c^2x^2+d^2y^2-1}{c^2(x^2-y^2)}\right)\ , 
\end{eqnarray}
where 
\beq
c=\frac{\sigma}{M}\ , \quad d=\frac{a}{M}\ , \quad c^2+d^2=1\ ,
\eeq
so that $\alpha=(\sigma-M)/a=(c-1)/d$.
Transition of this form of Kerr metric to the more familiar one associated with Boyer-Lindquist coordinates is accomplished by the map 
\beq
\label{trasftoBL}
t=t\ , \qquad
x=\frac{r-M}{\sigma}\ , \qquad 
y=\cos\theta\ , \qquad
\phi=\phi\ ,
\eeq
so that $x=1$ corresponds to the outer horizon $r=r_+=M+\sigma$.

\subsubsection{Erez-Rosen solution}

Similarly, for vanishing rotation parameter we recover the Erez-Rosen solution \cite{erez,novikov,young}. 
It is a solution of the static Weyl class of solutions (i.e. $\omega\equiv0$) with functions 
\beq
f_{ER}=\frac{x-1}{x+1}e^{-2qP_2Q_2}\ , \qquad 
\gamma_{ER}=\hat\gamma\ ,
\eeq
which reduce to the Schwarzschild solution
\beq
\label{schw}
f_S=\frac{x-1}{x+1}\ , \qquad
\gamma_S=\frac12\ln\left(\frac{x^2-1}{x^2-y^2}\right)
\eeq
when $q=0$ too.

\subsubsection{Hartle-Thorne solution}

The Hartle-Thorne solution is associated with the exterior field of a slowly rotating slightly deformed object \cite{ht67}. 
It is an approximate solution of the vacuum Einstein equations accurate to second order in the rotation parameter $a/M$ and to first order in the quadrupole parameter $q$, generalizing the Lense-Thirring spacetime \cite{lt18}.
The corresponding metric functions are given by 
\begin{eqnarray}\fl\qquad
f_{HT}&\simeq&\frac{x-1}{x+1}\left[1-q\left(2P_2Q_2-\ln\frac{x-1}{x+1}\right)\right]-\frac{x^2+x-2y^2}{(x+1)^3}\left(\frac{a}{M}\right)^2\ , \nonumber\\
\fl\qquad
\omega_{HT}&\simeq&2M\frac{1-y^2}{x-1}\left(\frac{a}{M}\right)\ , \nonumber\\
\fl\qquad
\gamma_{HT}&\simeq&\frac12 \ln \left(\frac{x^2-1}{x^2-y^2} \right)+2q (1-P_2)Q_1-\frac12\frac{1-y^2}{x^2-1}\left(\frac{a}{M}\right)^2\ ,
\end{eqnarray}
where terms of the order of $q(a/M)$ have also been neglected.

\section{Circular orbits on the symmetry plane}
\label{sec:cir}

Let us introduce the ZAMO family of fiducial observers, with four velocity
\beq
\label{n}
n=N^{-1}(\partial_t-N^{\phi}\partial_\phi)\ ,
\eeq
where $N=(-g^{tt})^{-1/2}$ and $N^{\phi}=g_{t\phi}/g_{\phi\phi}$ are the lapse and shift functions respectively. A suitable orthonormal frame adapted to  ZAMOs is given by
\beq
\label{zamoframe}
e_{\hat t}=n , \,\quad
e_{\hat x}=\frac1{\sqrt{g_{xx}}}\partial_x, \,\quad
e_{\hat y}=\frac1{\sqrt{g_{yy}}}\partial_y, \,\quad
e_{\hat \phi}=\frac1{\sqrt{g_{\phi \phi }}}\partial_\phi ,
\eeq
with dual
\beq\fl\quad
\omega^{{\hat t}}=N\rmd t\ , \quad \omega^{{\hat x}}=\sqrt{g_{xx}}\rmd x\ , \quad
\omega^{{\hat y}}= \sqrt{g_{yy}} \rmd y\ , \quad
\omega^{{\hat \phi}}=\sqrt{g_{\phi \phi }}(\rmd \phi+N^{\phi}\rmd t)\ .
\eeq

The 4-velocity $U$ of uniformly rotating circular orbits
can be parametrized either by the (constant) angular velocity with respect to infinity $\zeta$ or, equivalently, by the (constant) linear velocity  $\nu$ with respect to ZAMOs
\beq
\label{orbita}
U=\Gamma [\partial_t +\zeta \partial_\phi ]=\gamma [e_{\hat t} +\nu e_{\hat \phi}], \qquad \gamma=(1-\nu^2)^{-1/2}\ ,
\eeq
where $\Gamma$ is a normalization factor which assures that $U_\alpha U^\alpha =-1$  given by
\beq
\Gamma =\left[ N^2-g_{\phi\phi}(\zeta+N^{\phi})^2 \right]^{-1/2}=\frac{\gamma}{N}
\eeq
 and
\beq
\zeta=-N^{\phi}+\frac{N}{\sqrt{g_{\phi\phi}}} \nu .
\eeq
We limit our analysis to the motion on the symmetry plane ($y=0$) of the solution (\ref{metgen})--(\ref{metquev}), where there exists a large variety of special circular orbits \cite{bjdf,idcf1,idcf2,bjm}.

The prolate spheroidal coordinates in which the metric (\ref{metgen}) is written are adapted to the Killing symmetries of the spacetime itself and automatically select the family of static or \lq\lq threading'' observers, i.e. those at rest with respect to the coordinates, following the time coordinate lines. 
Threading observers have zero angular velocity, whereas their relative velocity with respect to ZAMOs is
\beq
\zeta_{\rm(thd)}=0\,, \qquad
\nu_{\rm(thd)}=\frac{f\omega}{\sigma\sqrt{x^2-1}}\ .
\eeq
ZAMOs are instead characterized by
\beq
\zeta_{\rm (zamo)}=-\frac{f^2\omega}{\sigma^2(x^2-1)-f^2\omega^2}\,, \qquad
\nu_{\rm(zamo)}=0\ .
\eeq

Co-rotating $(+)$ and counter-rotating $(-)$ timelike circular geodesics are characterized by the following linear velocities
\beq\fl\quad
\label{nugeo}
\nu_{({\rm geo})\, \pm}\equiv \nu_\pm =
\frac{fC\pm\left[f^2\omega^2-\sigma^2(x^2-1)\right]\sqrt{D}}{\sqrt{x^2-1}\sigma
\{f_x[f^2\omega^2+\sigma^2(x^2-1)]+2f(f^2\omega\omega_x-\sigma^2x)\}}
\ ,
\eeq
where
\begin{eqnarray}
C&=&-2\sigma^2(x^2-1)\omega f_x-f\{\omega_x[f^2\omega^2+\sigma^2(x^2-1)]-2\sigma^2x\omega\}\ , \nonumber\\
D&=&f^4\omega_x^2-\sigma^2f_x[f_x(x^2-1)-2xf]\ .
\end{eqnarray}
All quantities in the previous expressions are meant to be evaluated at $y=0$.
The corresponding timelike conditions $|\nu_\pm|<1$ together with the reality condition $D\geq0$ identify the
allowed regions for the \lq\lq radial'' coordinate where co/counter-rotating
geodesics exist. 
We refer to Ref. \cite{bglq} for a detailed discussion about the effect of the quadrupole moment on the causality condition.
There exists a finite range of values of $q$ wherein timelike circular geodesics are allowed: $q_1<q<q_3$ for co-rotating and $q_2<q<q_3$ for counter-rotating circular geodesics.
The critical values $q_1$, $q_2$ and $q_3$ of the quadrupole parameter can be (numerically) determined from Eq. (\ref{nugeo}).  
For instance, for a fixed distance parameter $x=4$ from the source and the choice of the rotation parameter $a/M=0.5$ we find $q_1\approx-105.59$, $q_2\approx-36.29$ and $q_3\approx87.68$.

\section{Tidal indicators}

We investigate here tidal forces, commonly associated with the Riemann tensor and more specifically with its electric and magnetic parts with respect to a generic timelike congruence. Let us denote by $u$ the corresponding unit tangent vector.
 
The electric and magnetic parts of the Weyl tensor $C_{\alpha\beta\gamma\delta}$ with respect to a generic timelike congruence with unit tangent vector $u$ are defined as \cite{book}
\beq
\label{weylsplitu}
E(u)_{\alpha\beta}=C_{\alpha\mu\beta\nu}u^\mu u^\nu\,,\qquad 
H(u)_{\alpha\beta}=-C^*_{\alpha\mu\beta\nu}u^\mu u^\nu\,.
\eeq 
These spatial fields are both symmetric and tracefree.
The electric part is associated with tidal gravity, whereas the magnetic part describes differential dragging of inertial frames. 
Some tools for visualizing the spacetime curvature through the electric and magnetic parts of the Weyl tensor have been recently introduced in Ref. \cite{owenetal}, where the nonlinear dynamics of curved spacetime in merging black hole binaries has been investigated by using numerical simulations. 
We recall that in a vacuum spacetime, which is just the case we are considering, the Weyl and Riemann tensors coincide.

The simplest way to built up scalar quantities through the electric and magnetic parts of the Riemann tensor which are representative of them and serve as \lq\lq tidal indicators" in the study of tidal effects is to take the trace of their square.
One can then consider the following electric and magnetic tidal indicators \cite{kerr_tidal}
\beq
\label{tidalind}
{\mathcal T}_E(u)= {\rm Tr}[E(u)^2]\,,\qquad 
{\mathcal T}_H(u)= {\rm Tr}[H(u)^2]\,.
\eeq
They are related to the curvature tensor as well as to the particle/observer undergoing tidal deformations.
One could also consider other more involved tidal invariants constructed from the covariant derivative of the curvature tensor.
Such invariants have received some attention in the recent literature in order to investigate both geometrical and topological properties of certain classes of static as well as stationary spacetimes (see, e.g., Refs. \cite{lake,mukherjee,saa}). 
However, in the context of tidal problems differential invariants are of interest only when using Fermi coordinate tidal potential, as discussed in Ref. \cite{mino}.
We will not address this problem in the present paper. 

Let $u=n$ be the unit tangent vector to the ZAMO family of observer given by Eq. (\ref{n}) with adapted frame (\ref{zamoframe}).
The relevant nonvanishing frame components of the electric and magnetic parts of the Riemann tensor are given by $E(n)_{11}$, $E(n)_{33}$ and $H(n)_{12}$
with
\beq
E(n)_{11}+E(n)_{22}=-E(n)_{33}\,.
\eeq
They are listed in Appendix A.
The tidal indicators (\ref{tidalind}) then  turn out to be given by
\begin{eqnarray}
\label{tidalindzamo}
{\mathcal T}_E(n)&=&2\{[E(n)_{11}]^2+[E(n)_{22}]^2+E(n)_{11}E(n)_{22}\}\,,\nonumber\\
{\mathcal T}_H(n)&=&2[H(n)_{12}]^2\,.
\end{eqnarray}

Let now $U$ be tangent to a uniformly rotating timelike circular orbit on the symmetry plane.
We find
\footnote{These relations were first derived in Ref. \cite{kerr_tidal}.
Note that the last line of Eq. (4.5) there was misprinted by an overall minus sign, corrected here.
Such a misprint also affected the equivalent form (4.6), which was but never used.}
\begin{eqnarray}\fl\qquad
\label{tidalind2}
{\mathcal T}_E(U)&=&\gamma^4\left\{
{\mathcal T}_E(n)(\nu^4+1)-4H(n)_{12}(E(n)_{11}-E(n)_{22})\nu(\nu^2+1)\right.\nonumber\\
\fl\quad
&&\left.-2\nu^2([E(n)_{11}]^2+[E(n)_{22}]^2+4E(n)_{11}E(n)_{22}-4[H(n)_{12}]^2)
\right\}\,,\nonumber\\
\fl\qquad
{\mathcal T}_H(U)&=& {\mathcal T}_H(n) \gamma^4(\nu-\nu_*)^2(\nu-\bar\nu_*)^2\,,
\end{eqnarray}
where 
\beq
\label{nureldef}
\nu_*=W-\sqrt{W^2-1}
\,, \qquad
W=\frac{E(n)_{11}-E(n)_{22}}{2H(n)_{12}}
\,,
\eeq
and we have used the notation $\bar\nu_*=1/\nu_*$.
After expliciting the Lorentz factor $\gamma^4=1/(1-\nu^2)^2$ and rearranging terms we can derive from Eq. (\ref{tidalind2}) the following relation
\beq
\label{relTEHinv}
{\mathcal T}_E(U)={\mathcal T}_E(n)-{\mathcal T}_H(n)+{\mathcal T}_H(U)\,.
\eeq
This is actually an invariance relation also involving the curvature invariants.
In fact, it is possible to show that
\beq
\label{relTEH}
{\mathcal T}_E(U)-{\mathcal T}_H(U)=\frac{K}{8}={\mathcal T}_E(n)-{\mathcal T}_H(n)\,,
\eeq
where $K$ is the Kretschmann invariant of the spacetime (evaluated on the equatorial plane $y=0$, see Appendix A), i.e.,
\beq
\label{Kdef}
K=C_{\alpha\beta\gamma\delta}C^{\alpha\beta\gamma\delta}\big|_{y=0}\,.
\eeq
Its behavior as a function of the distance parameter is shown in Fig. \ref{fig:1}(a) for a fixed value of the rotation parameter and different values of the quadrupole parameter.
As a consequence, since $K$ does not depend on $\nu$, all along the family of circular orbits parametrized by $\nu$ both tidal indicators have their extremal values simultaneously, i.e., 
\beq
\frac{d {\mathcal T}_E(U)}{d\nu}=\frac{d {\mathcal T}_H(U)}{d\nu}\,.
\eeq

\subsection{Super-energy density and super-Poynting vector}

Bel \cite{bel58} and Robinson \cite{robinson} first introduced the Bel-Robinson super-energy-momentum tensor for the gravitational field in vacuum in terms of the Weyl curvature tensor in analogy with electromagnetism (see also Refs. \cite{maartens98,Bonilla}) 
\beq\label{BelR}
 T_{\alpha\beta} {}^{\gamma\delta} 
 = \frac{1}{2}
   ( C_{\alpha\rho\beta\sigma} C^{\gamma\rho\delta\sigma}
   + {}^* C_{\alpha\rho\beta\sigma} {}^* C^{\gamma\rho\delta\sigma} )
\ .
\eeq
The super-energy density and the super-Poynting vector associated with a generic observer $u$ are given by
\begin{eqnarray}
 {\mathcal E}^{(\rm{g})}(u) 
 &=& T_{\alpha\beta\gamma\delta} u^\alpha u^\beta u^\gamma u^\delta
 =\frac{1}{2} [ {\mathcal T}_E(u)+{\mathcal T}_H(u) ]\ ,\nonumber\\
 P^{(\rm{g})} (u)_\alpha  
 &=& T_{\alpha\beta\gamma\delta} u^\beta u^\gamma u^\delta
= [ E(u)\times_u H(u)]_\alpha\ ,
\end{eqnarray}
where 
\beq
[ E(u)\times_u H(u)]_{\alpha} 
= \eta (u)_{\alpha\beta\gamma}
   E(u)^{\beta} {}_{\delta}\, H(u)^{\delta\gamma}\ 
\eeq
and the spatial unit-volume $3$-form has been introduced, i.e.,
\beq
\eta(u)_{\alpha\beta\gamma}=u^\mu \eta_{\mu\alpha\beta\gamma}\,.
\eeq

Bel showed that for Petrov types I and D, an observer always exists for which the super-Poynting vector vanishes: this observer aligns the electric and magnetic parts of the Weyl tensor in the sense that they are both diagonalized and therefore commute. 
For black hole spacetimes, the Carter's observer family plays this role at each spacetime point.
The same property is also shared by the observers $\nu_*$ in this more general context.
In fact, the super-Poynting vector for a circularly rotating orbit $U$ turns out to have a single nonvanishing frame component along $e_{\hat \phi}$ given by
\beq\fl\quad
P^{(\rm{g})} (U)_{\hat \phi}= \frac{\gamma^4}{2\nu_*}{\mathcal T}_H(n)(\nu-\nu_*)(\nu-\bar\nu_*)[(1+\nu_*^2)(1+\nu^2)-4\nu\nu_*]\,.
\eeq
Furthermore, the orbit associated with $\nu_*$ is also characterized by the following relation involving the super-energy density (which can be easily evaluated from Eq. (\ref{tidalind2}))
\beq
\frac{d {\mathcal E}^{(\rm{g})}(U)}{d\nu}=-4\gamma^2P^{(\rm{g})} (U)_{\hat \phi}\,,
\eeq
or in terms of the rapidity parameter $\alpha$ such that $\nu=\tanh\alpha$
\beq
\frac{d {\mathcal E}^{(\rm{g})}(U)}{d\alpha}=-4P^{(\rm{g})} (U)_{\hat \phi}\,,
\eeq
so remembering the structure of Hamilton's equations for conjugate variables.
The observers $\nu_*$ thus correspond to vanishing super-Poynting vector and minimal super-energy density.

\subsection{Discussion}

Eq. (\ref{tidalind2}) implies that ${\mathcal T}_H(U)$ vanishes for $\nu=\nu_*$, regardless of the value of the quadrupole parameter $q$.
The family of observers identified by this 4-velocity plays the role of Carter's family in such a generalized Kerr spacetime.
In the Kerr case in Boyer-Lindquist coordinates Carter's observer velocity is given by
\beq
\nu_{\rm (car)}=\frac{a\sqrt{\Delta}}{r^2+a^2}\,.
\eeq
The main property of Carter's observers world lines is to be the unique timelike world lines belonging to the intersection of the Killing two-plane ($t,\,\phi$) with the two-plane spanned by the two independent principal null directions of the Kerr spacetime. 
In contrast, the QM solution is Petrov type I with four independent principal null directions and hence the above property is lost.

\begin{figure} 
\begin{center}
\includegraphics[scale=.35]{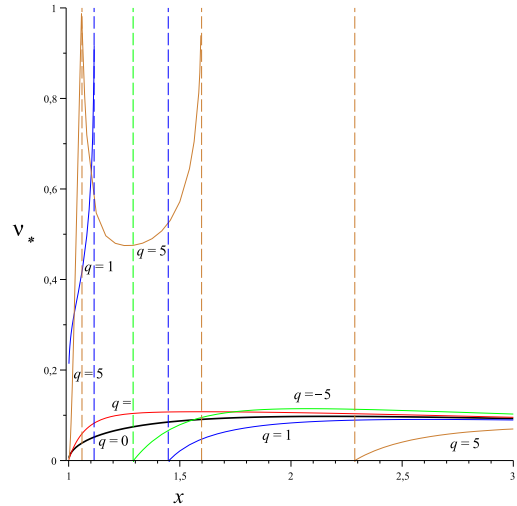}
\end{center}
\caption{
The behavior of the linear velocity $\nu_*$ as a function of the distance parameter is shown for the choice $a/M=0.5$ and different values of the quadrupole $q=[-5,-1,0,1,5]$.
The thick black solid curve refers to the Kerr case ($q=0$), i.e., to the Carter's 4-velocity.
Dashed vertical lines correspond to observer horizons.
} 
\label{fig:2}
\end{figure}

Fig. \ref{fig:2} shows how the behavior of the generalized Carter's observer velocity $\nu_*$ as a function of the distance parameter modifies due to the presence of the quadrupole. 
Carter's observers are defined outside the outer event horizon of the Kerr spacetime, corresponding to $x=1$.
For negative values of $q$, the observer horizon associated with such generalized Carter's observers is located at a certain value $x>1$, which increases as $q$ becomes increasingly negative. 
A similar situation occurs also for positive values of $q$, but additional allowed branches also appear in the inner region.
 
\begin{figure} 
\typeout{*** EPS figure 3}
\begin{center}
$\begin{array}{cc}
\includegraphics[scale=0.28]{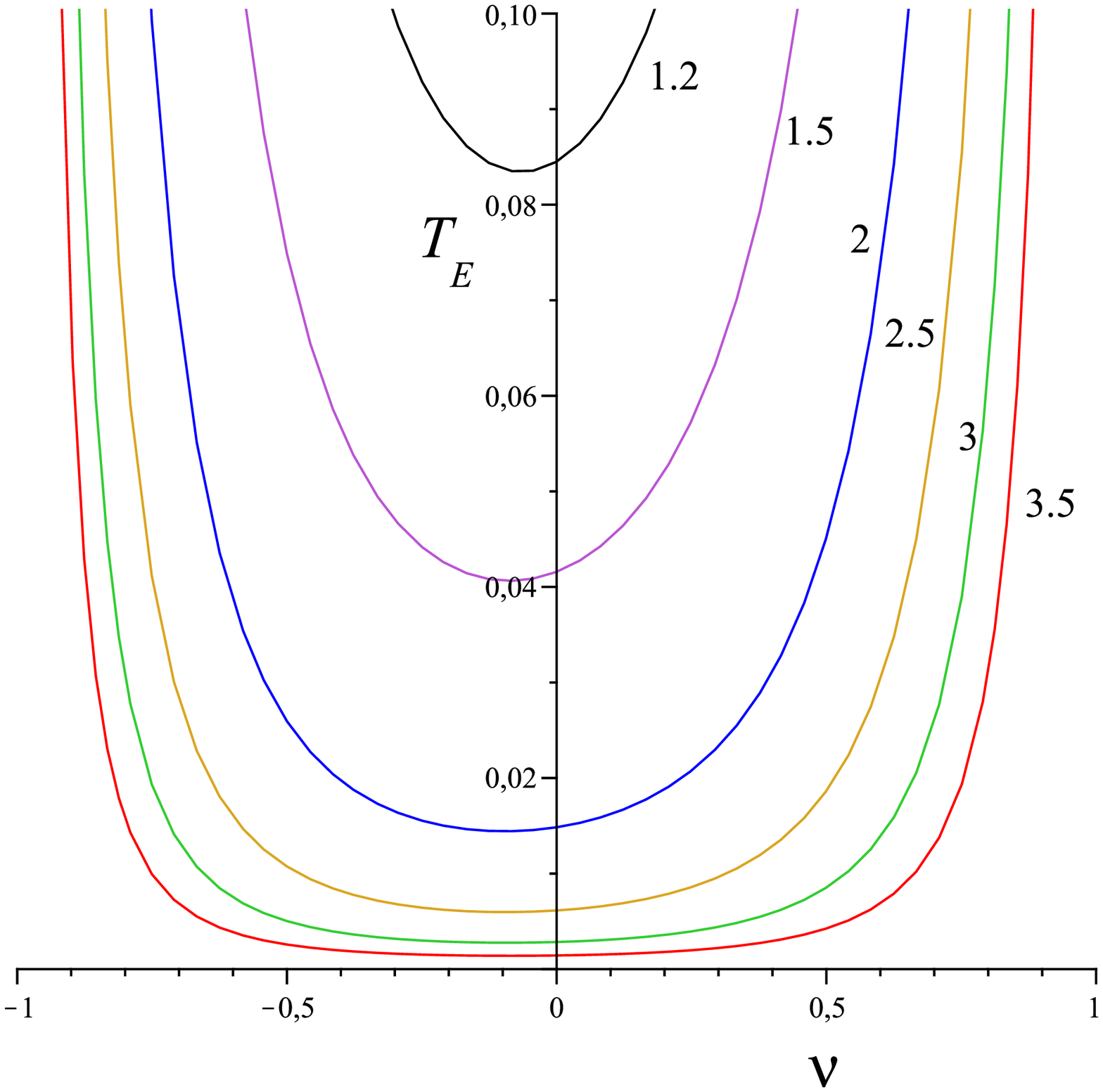}&\quad
\includegraphics[scale=0.28]{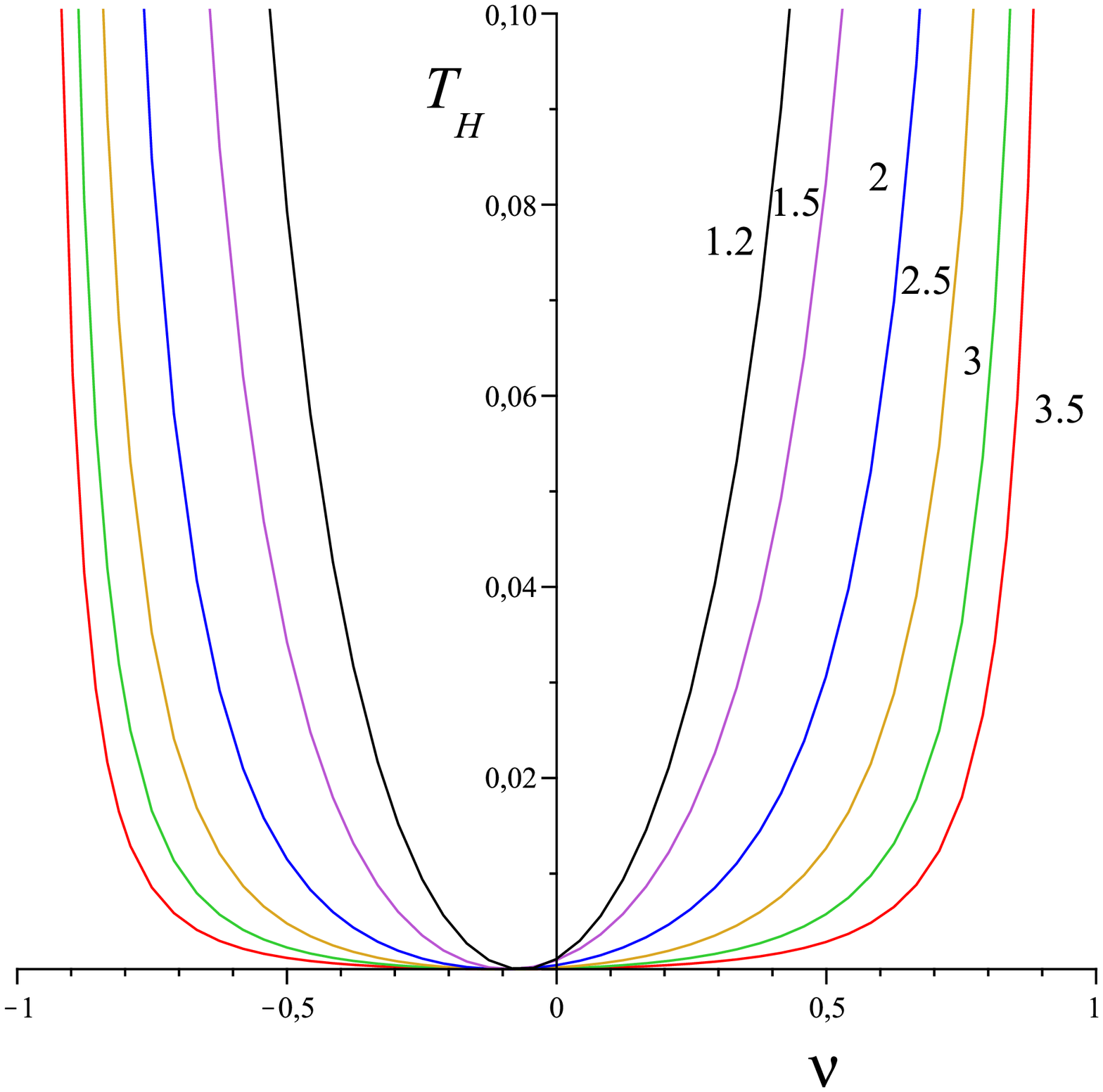}\\[.4cm]
\quad\mbox{(a)}\quad &\quad \mbox{(b)}\\[.6cm]
\includegraphics[scale=0.28]{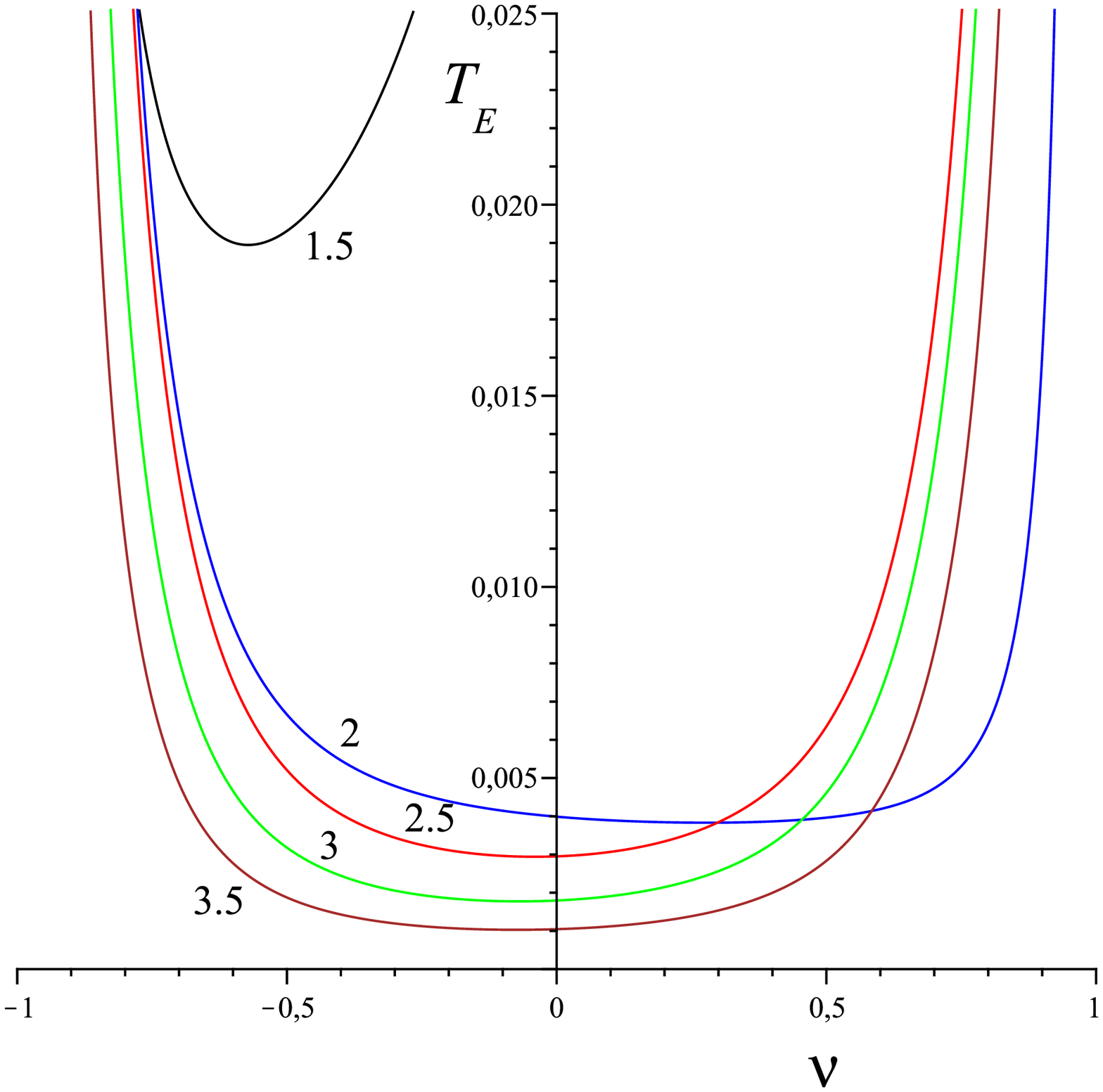}&\quad
\includegraphics[scale=0.28]{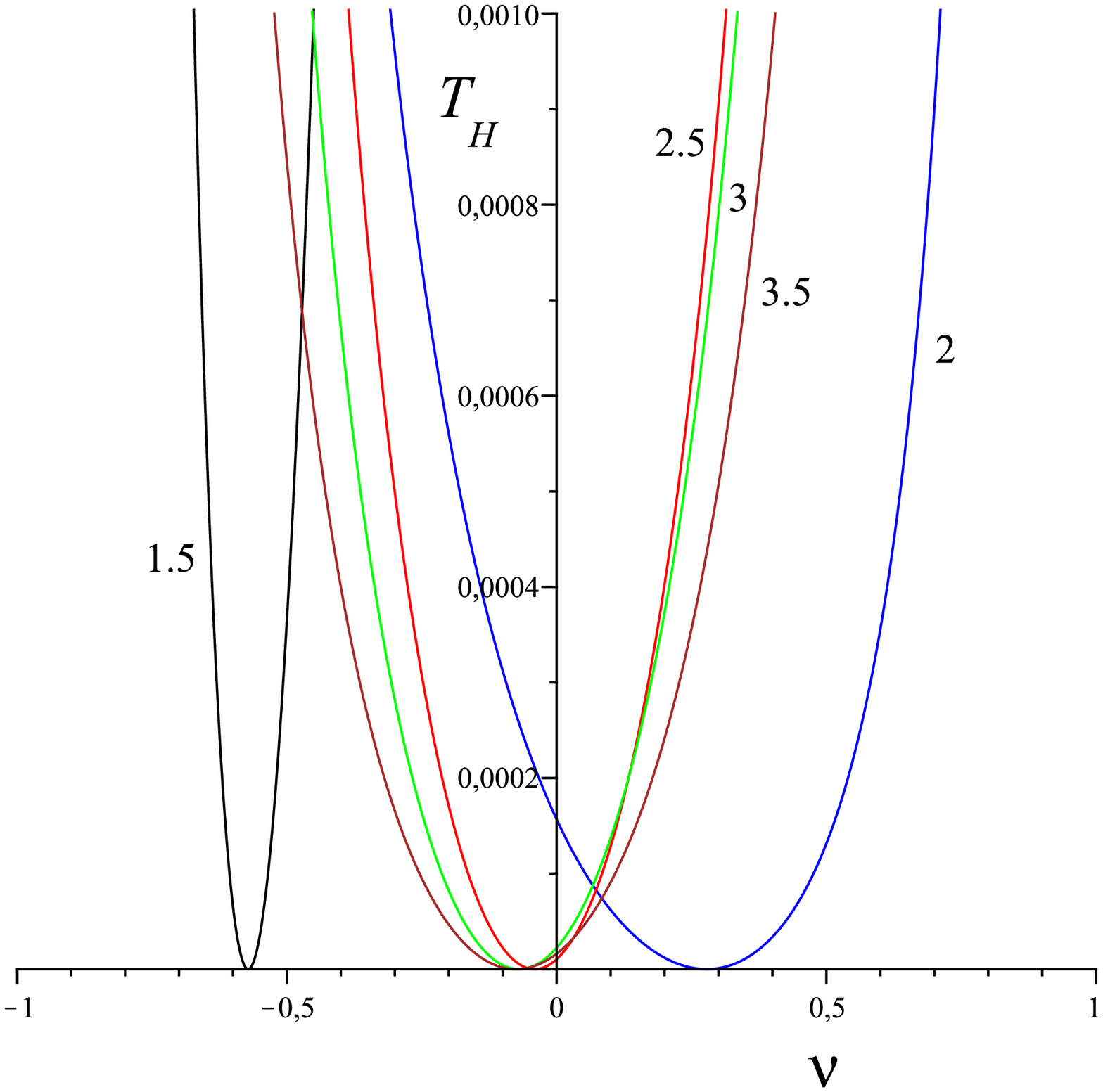}\\[.4cm]
\quad\mbox{(c)}\quad &\quad \mbox{(d)}\\
\end{array}$
\end{center}
\caption{The behaviors of the tidal indicators ${\mathcal T}_E(U)$ and ${\mathcal T}_H(U)$ as functions of $\nu$ are shown in panels (a), (c) for $q=0$ and (b), (d) for $q=5$ respectively for the choice $a/M=0.5$ and different values of the coordinate $x=[1.2,1.5,2, 2.5, 3, 3.5]$. 
For negative values of $q$ the behaviors are very similar to the Kerr case.
For $q$ increasingly negative the values of ${\mathcal T}_E(U)$ at the minimum of each curve increase, whereas the curves corresponding to ${\mathcal T}_H(U)$ shrink to the vertical axis. 
Units on the vertical axis are chosen so that $M=1$.   
}
\label{fig:3}
\end{figure}

\begin{figure}
\begin{center}
$\begin{array}{cc}
\includegraphics[scale=.28]{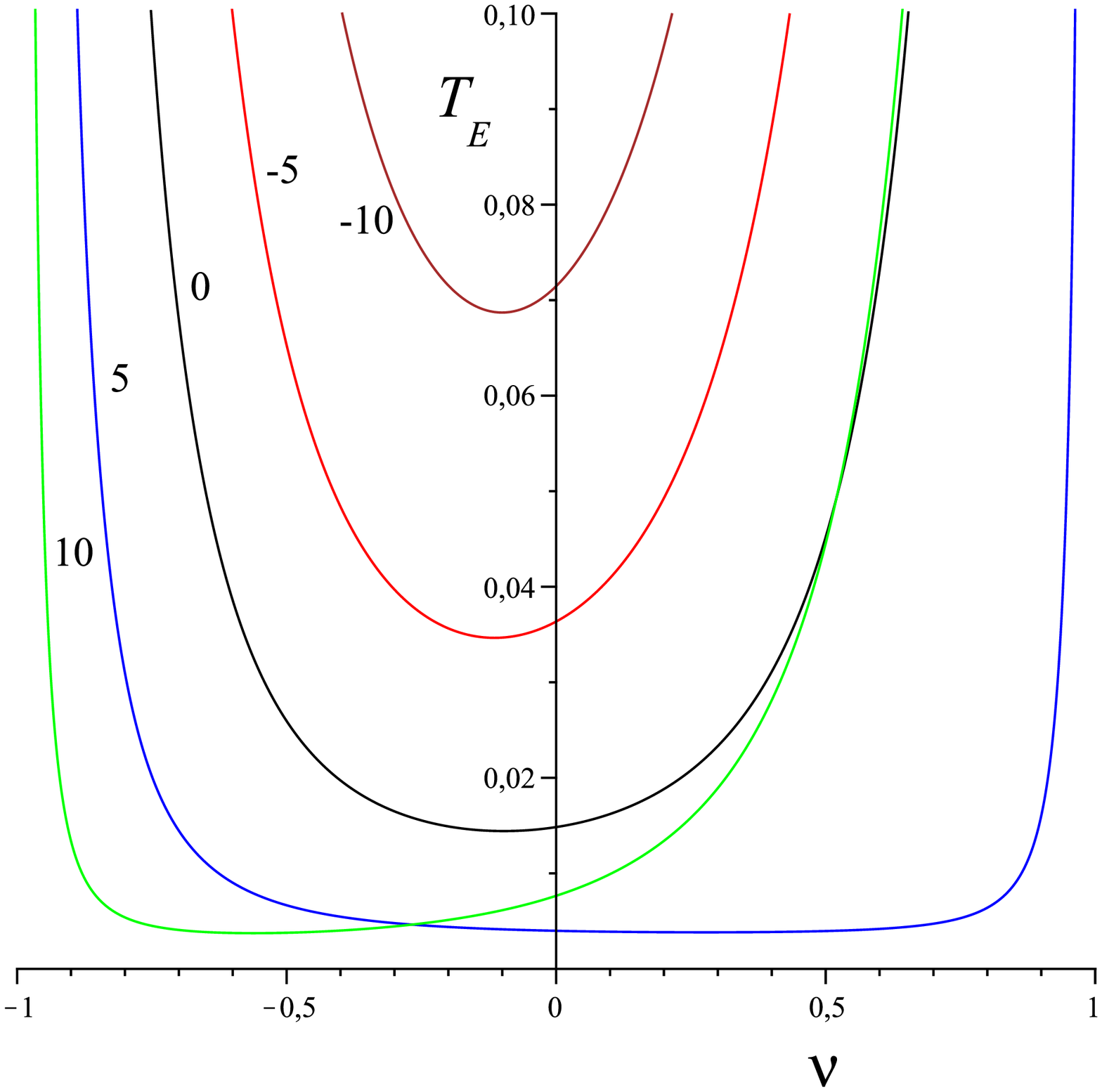}& 
\includegraphics[scale=.28]{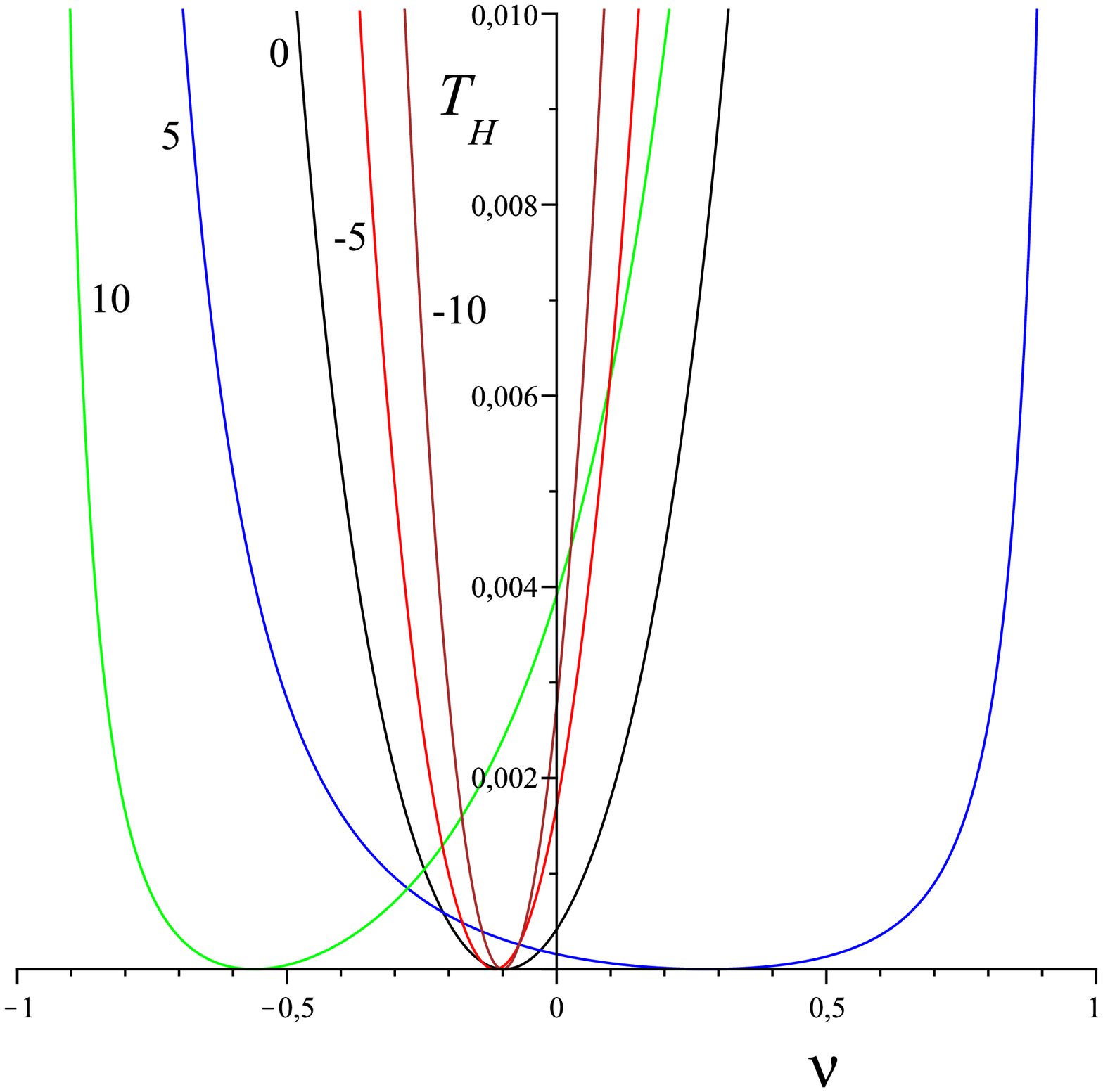}\\[0.4cm] 
\mbox{(a)}&\mbox{(b)}\\
\end{array}$
\end{center}
\caption {The behaviors of the tidal indicators ${\mathcal T}_E(U)$ and ${\mathcal T}_H(U)$ as functions of $\nu$ are shown in panels (a) and (b) respectively for the choice $a/M=0.5$, with a fixed value $x=2$ of the distance parameter and different values of $q=[-10,-5, 0, 5, 10]$. 
Units on the vertical axis are chosen so that $M=1$.
} 
\label{fig:4}
\end{figure}

In contrast with the Kerr case where Carter's observers are the only ones which measure zero magnetic tidal indicator, this property is shared here also by other observer families for special values of $q$.
This is evident from Figs. \ref{fig:3} and \ref{fig:4}, where the behaviors of the tidal indicators as functions of $\nu$ are shown for a fixed value of the quadrupole and different values of the distance parameter in comparison with the Kerr case (Fig. \ref{fig:3}) and for a fixed value of $x$ and different values of $q$ (Fig. \ref{fig:4}). 
In fact, we see that ${\mathcal T}_H(U)$ vanishes many times for different values of $\nu$ corresponding either to  different $x$ with fixed $q$, or to different $q$ with fixed $x$. 

\begin{figure} 
\typeout{*** EPS figure 1}
\begin{center}
$\begin{array}{cc}
\includegraphics[scale=0.28]{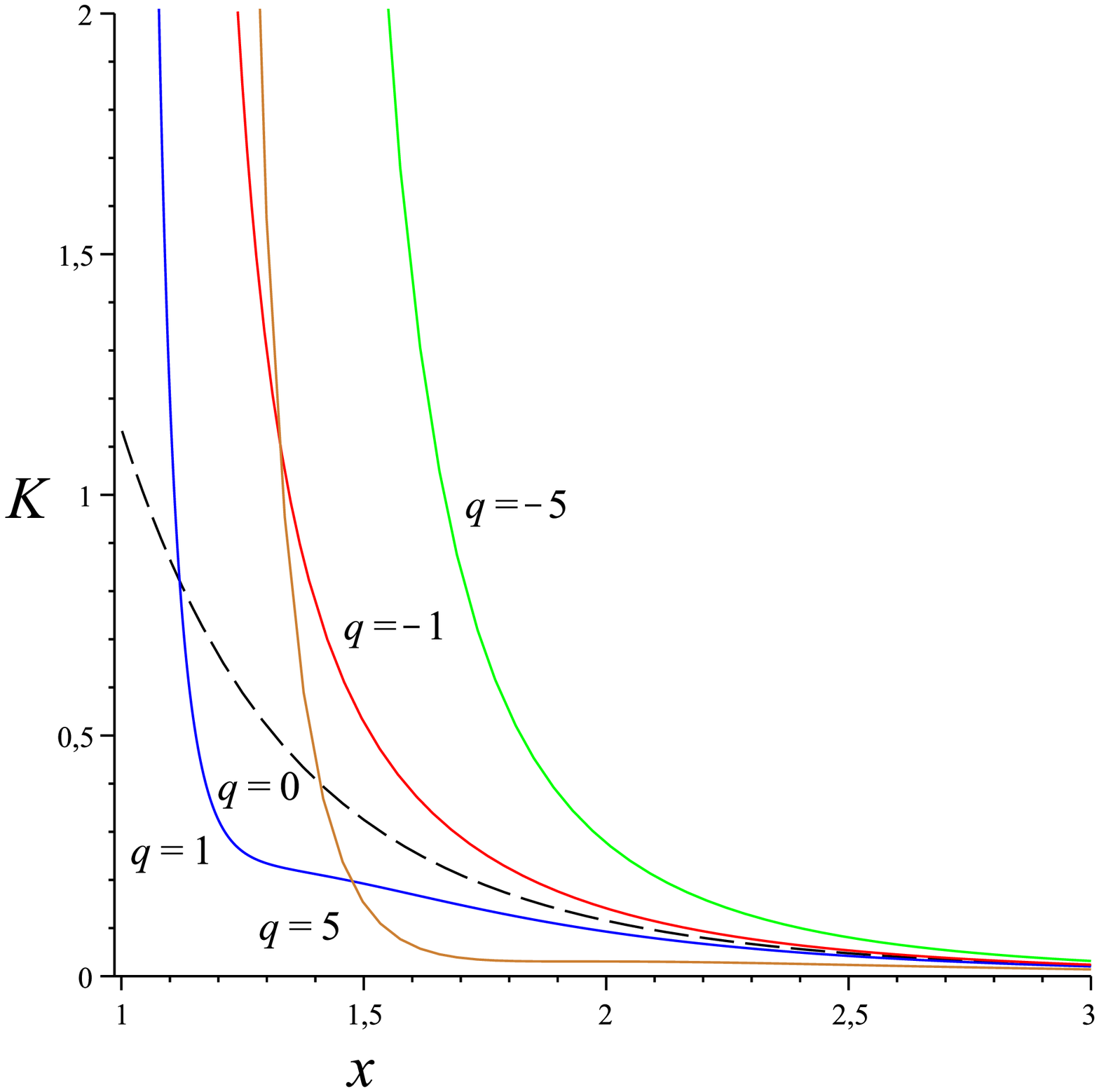}&\quad
\includegraphics[scale=0.28]{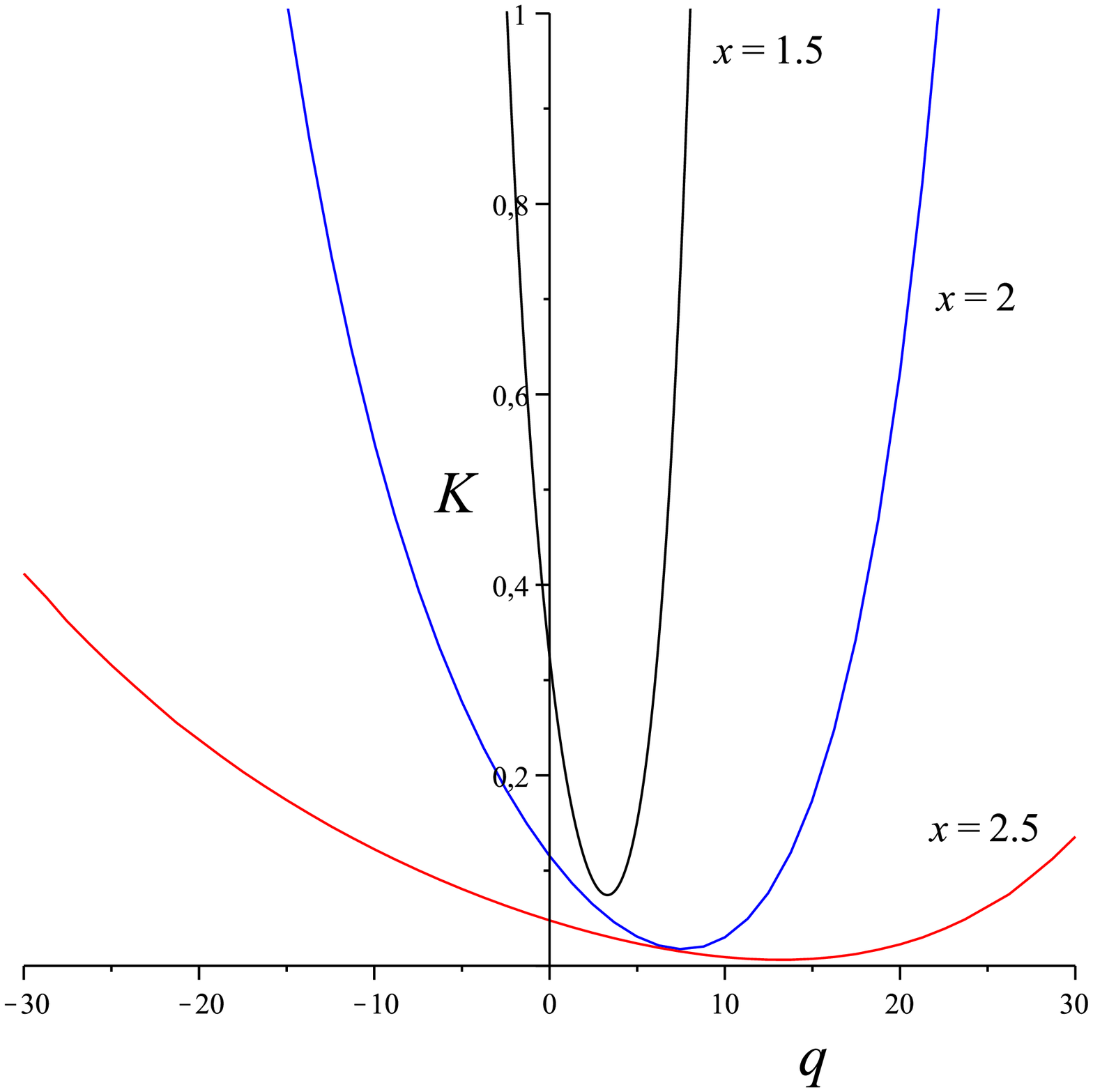}\\[.4cm]
\quad\mbox{(a)}\quad &\quad \mbox{(b)}\\
\end{array}$
\end{center}
\caption{The behavior of the Kretchmann invariant $K$ evaluated on the symmetry plane is shown as a function of the distance parameter for the choice $a/M=0.5$ and different values of $q=[-5,-1,0,1,5]$ in panel (a).
Panel (b) shows instead its behavior as a function of $q$ for different values of $x=[1.5,2,2.5]$
Units on the vertical axis are chosen so that $M=1$.   
}
\label{fig:1}
\end{figure}

\begin{figure} 
\typeout{*** EPS figure 5}
\begin{center}
$\begin{array}{cc}
\includegraphics[scale=0.28]{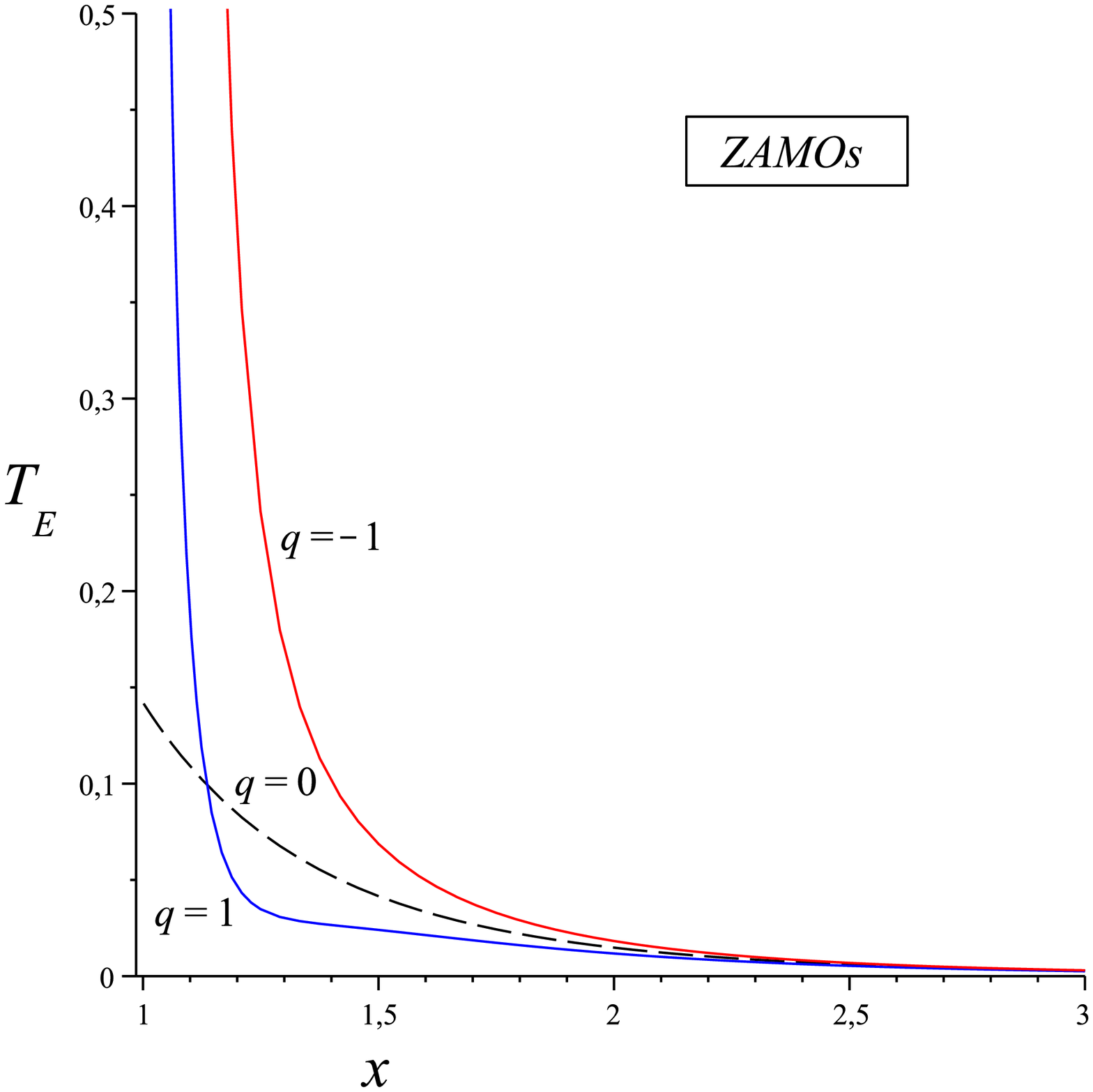}&\quad
\includegraphics[scale=0.28]{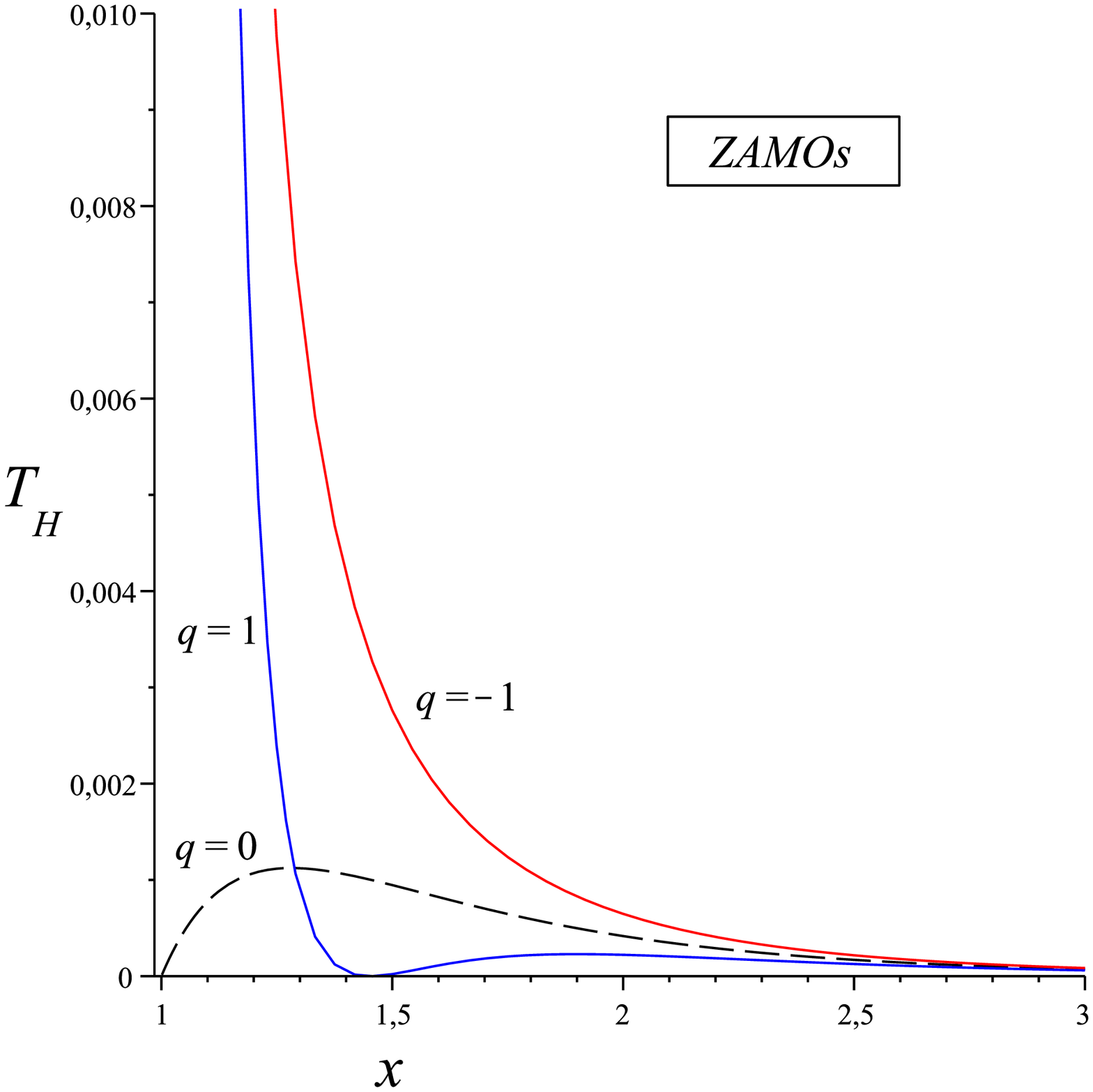}\\[.4cm]
\quad\mbox{(a)}\quad &\quad \mbox{(b)}\\
\end{array}$
\end{center}
\caption{The behaviors of the tidal indicators ${\mathcal T}_E(U)$ and ${\mathcal T}_H(U)$ as measured by ZAMOs are shown as functions of the distance parameter for the choice $a/M=0.5$ and different values of $q=[-1,0,1]$.
Dashed curves correspond to the Kerr case (i.e., $q=0$).
Units on the vertical axis are chosen so that $M=1$.   
}
\label{fig:5}
\end{figure}

Correspondingly the electric type tidal indicator takes its minimum value, which can be further reduced by suitably choosing the quadrupole parameter, but cannot be made vanishing.
In fact, the invariant relation (\ref{relTEH}) implies for instance that ${\mathcal T}_E(U)$ is minimum when ${\mathcal T}_H(U)=0$ and $K$ is mimimum as well.
Fig. \ref{fig:1}(b) shows the behavior of $K$ as a function of $q$ for different values of the distance parameter.
$K$ reaches an absolute minimum only for positive values of $q$, which increase for increasing $x$.   
The behaviors of the tidal indicators as measured by ZAMOs, static and geodesic observers are shown in Figs. \ref{fig:3}--\ref{fig:5}, respectively, as functions of the distance parameter for different values of $q$.
The behavior is Kerr-like for negative values of $q$.
For positive values of $q$ instead the situation significantly modifies with respect to the Kerr case, the magnetic tidal indicator vanishing many times and correspondingly the electric tidal indicator showing a damped oscillating behavior.


\begin{figure} 
\typeout{*** EPS figure 6}
\begin{center}
$\begin{array}{cc}
\includegraphics[scale=0.28]{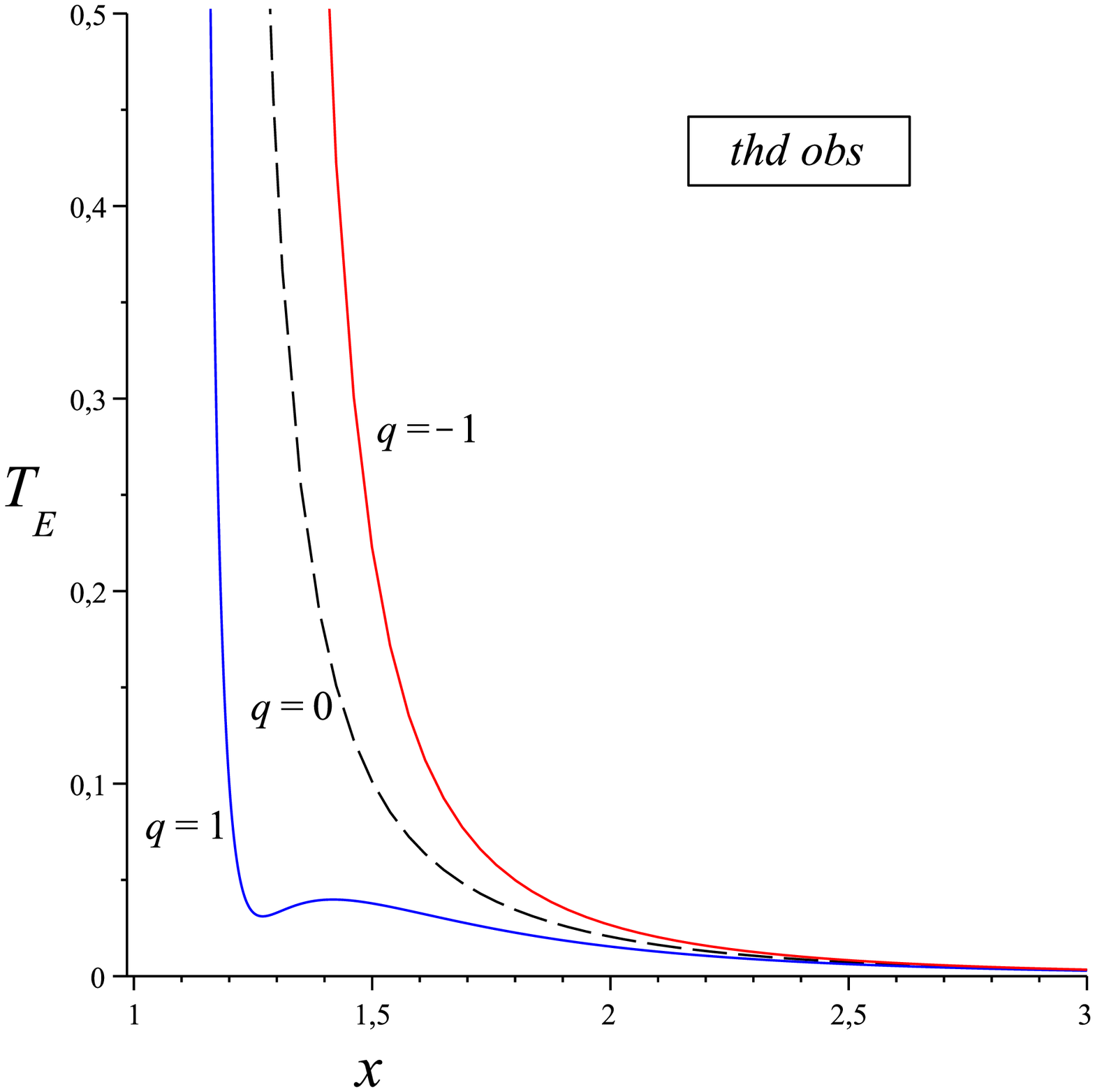}&\quad
\includegraphics[scale=0.28]{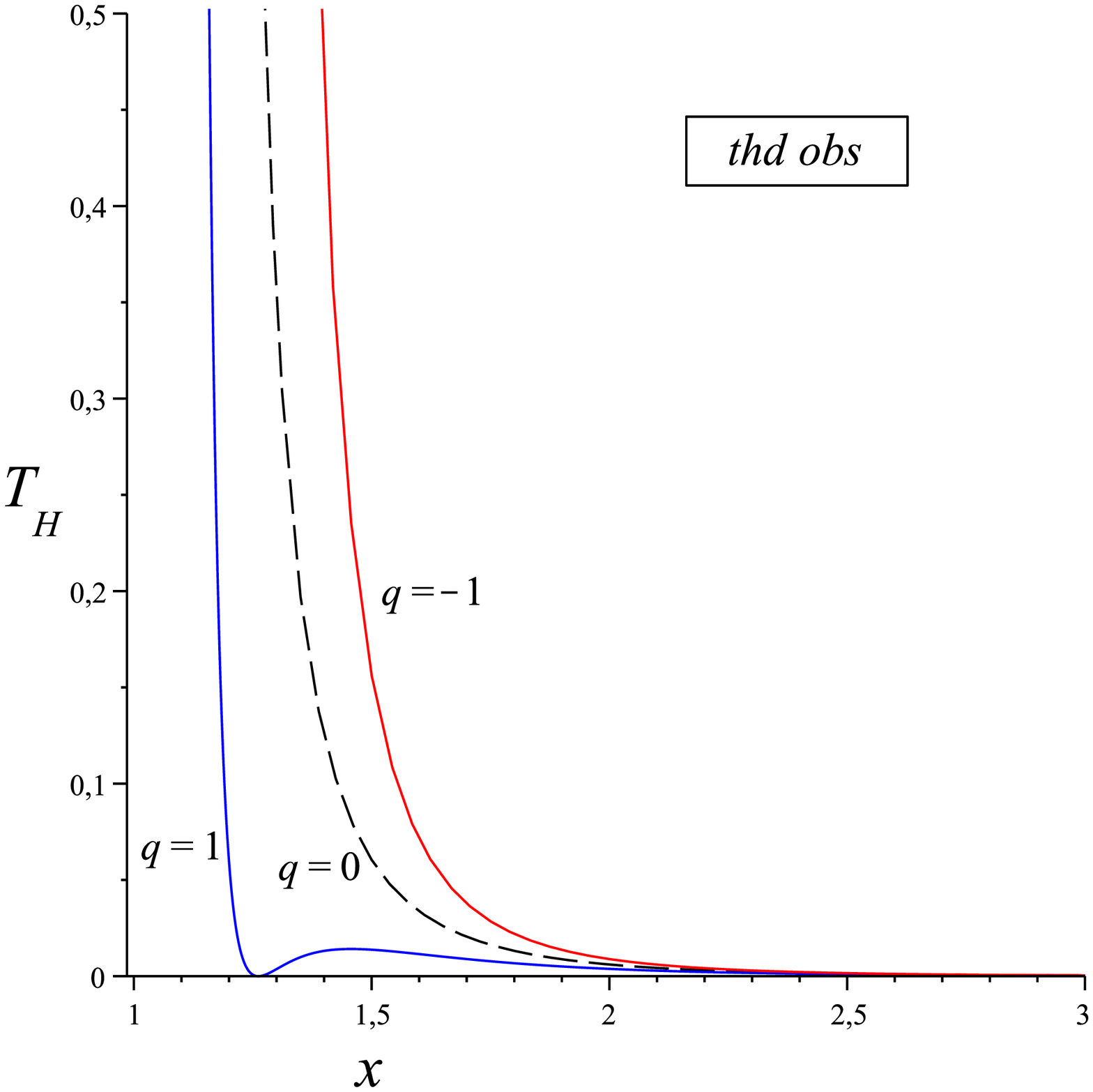}\\[.4cm]
\quad\mbox{(a)}\quad &\quad \mbox{(b)}\\
\end{array}$
\end{center}
\caption{The behaviors of the tidal indicators ${\mathcal T}_E(U)$ and ${\mathcal T}_H(U)$ as measured by static observers are shown as functions of the distance parameter for the choice $a/M=0.5$ and different values of $q=[-1,0,1]$.
Dashed curves correspond to the Kerr case (i.e., $q=0$).
Units on the vertical axis are chosen so that $M=1$.   
}
\label{fig:6}
\end{figure}


\begin{figure} 
\typeout{*** EPS figure 7}
\begin{center}
$\begin{array}{cc}
\includegraphics[scale=0.28]{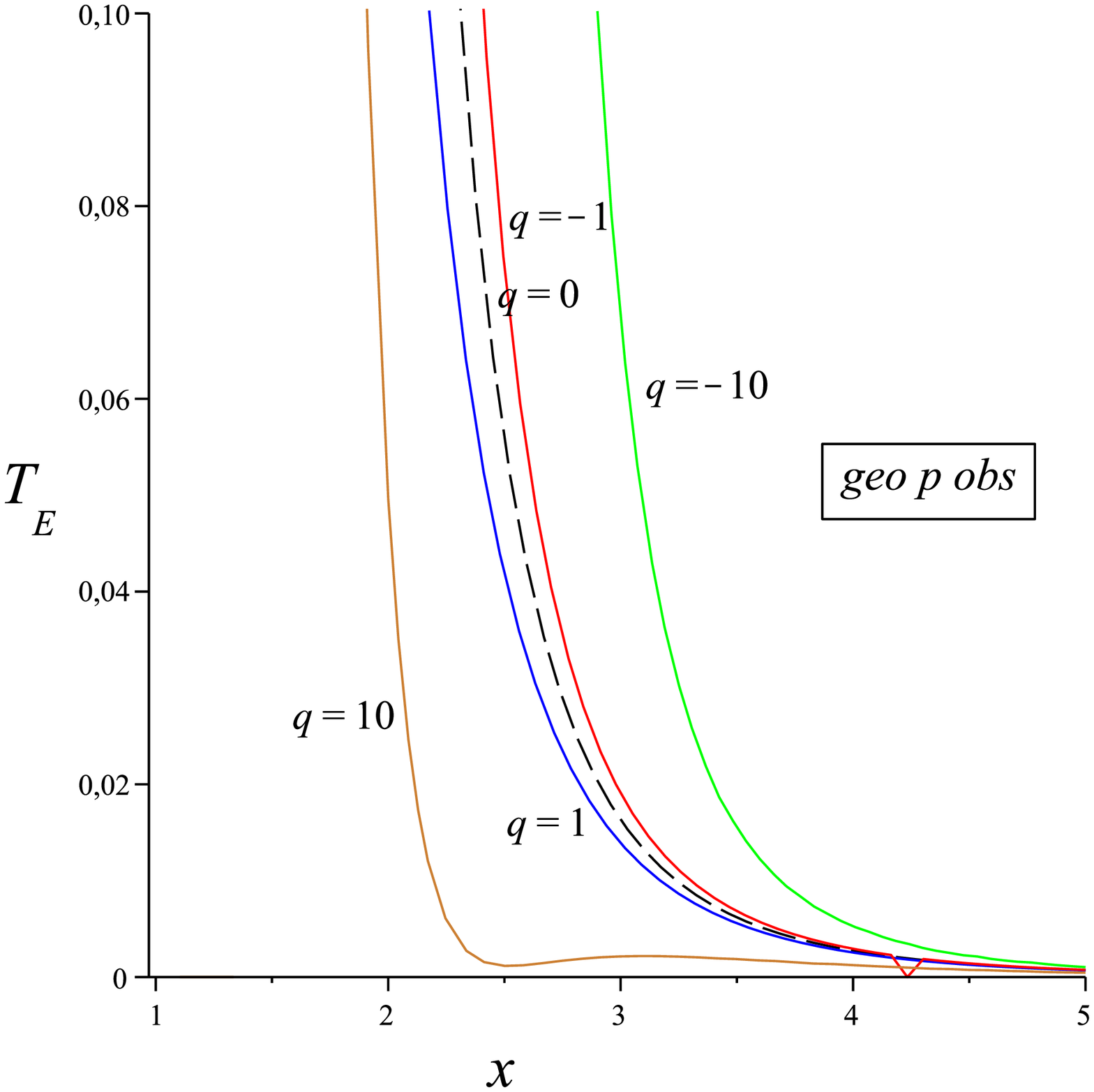}&\quad
\includegraphics[scale=0.28]{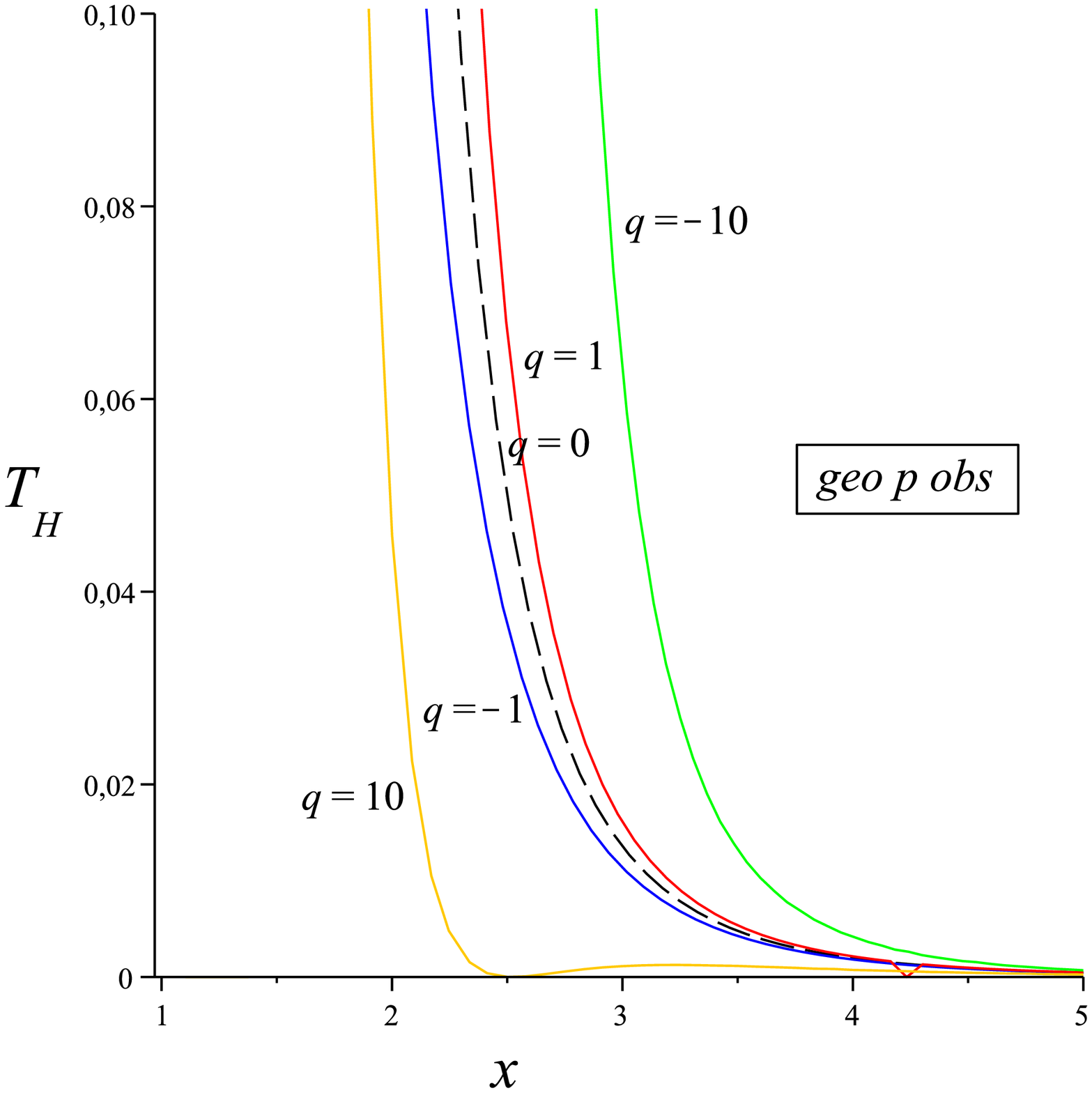}\\[.4cm]
\quad\mbox{(a)}\quad &\quad \mbox{(b)}\\[.6cm]
\includegraphics[scale=0.28]{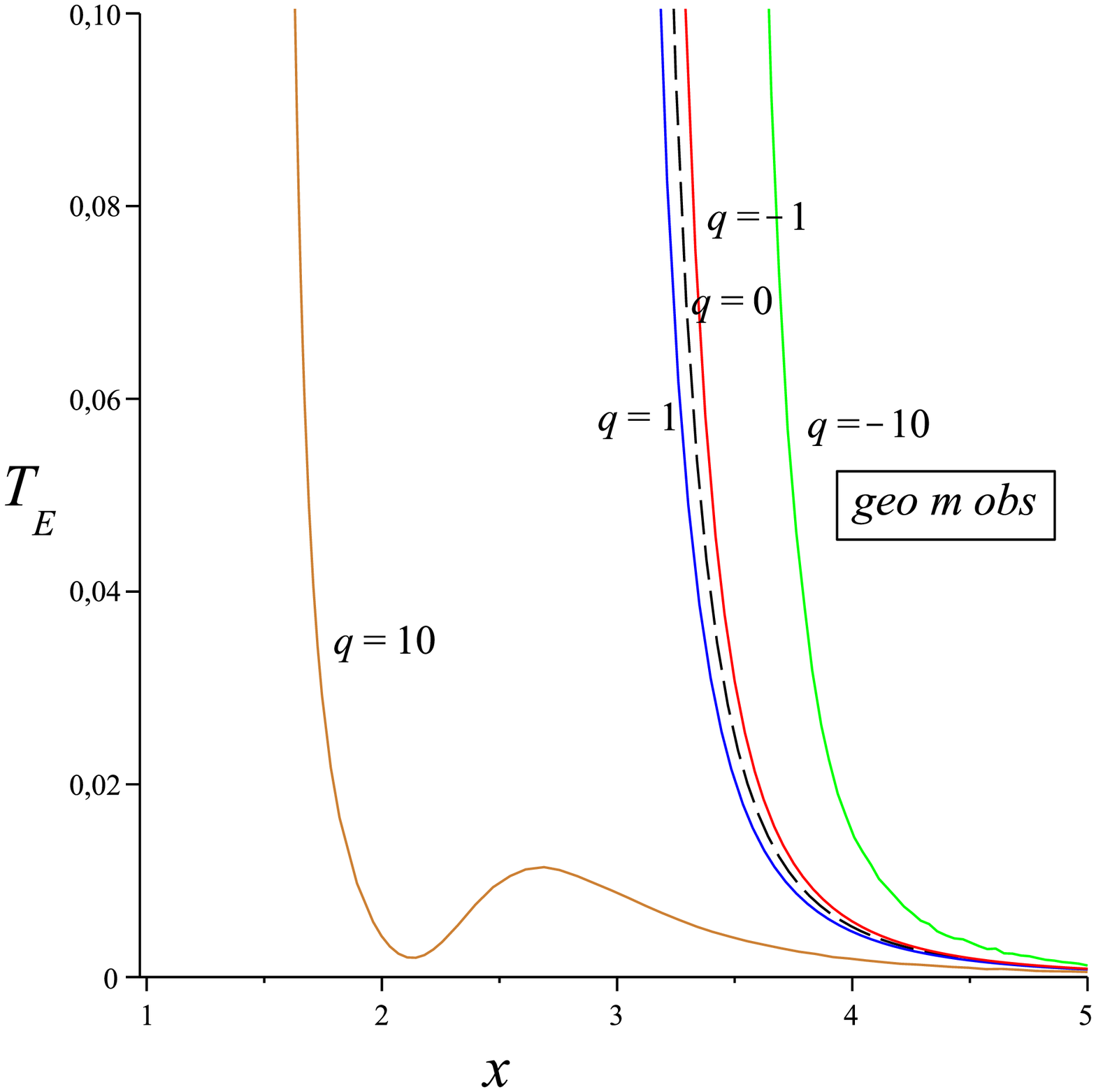}&\quad
\includegraphics[scale=0.28]{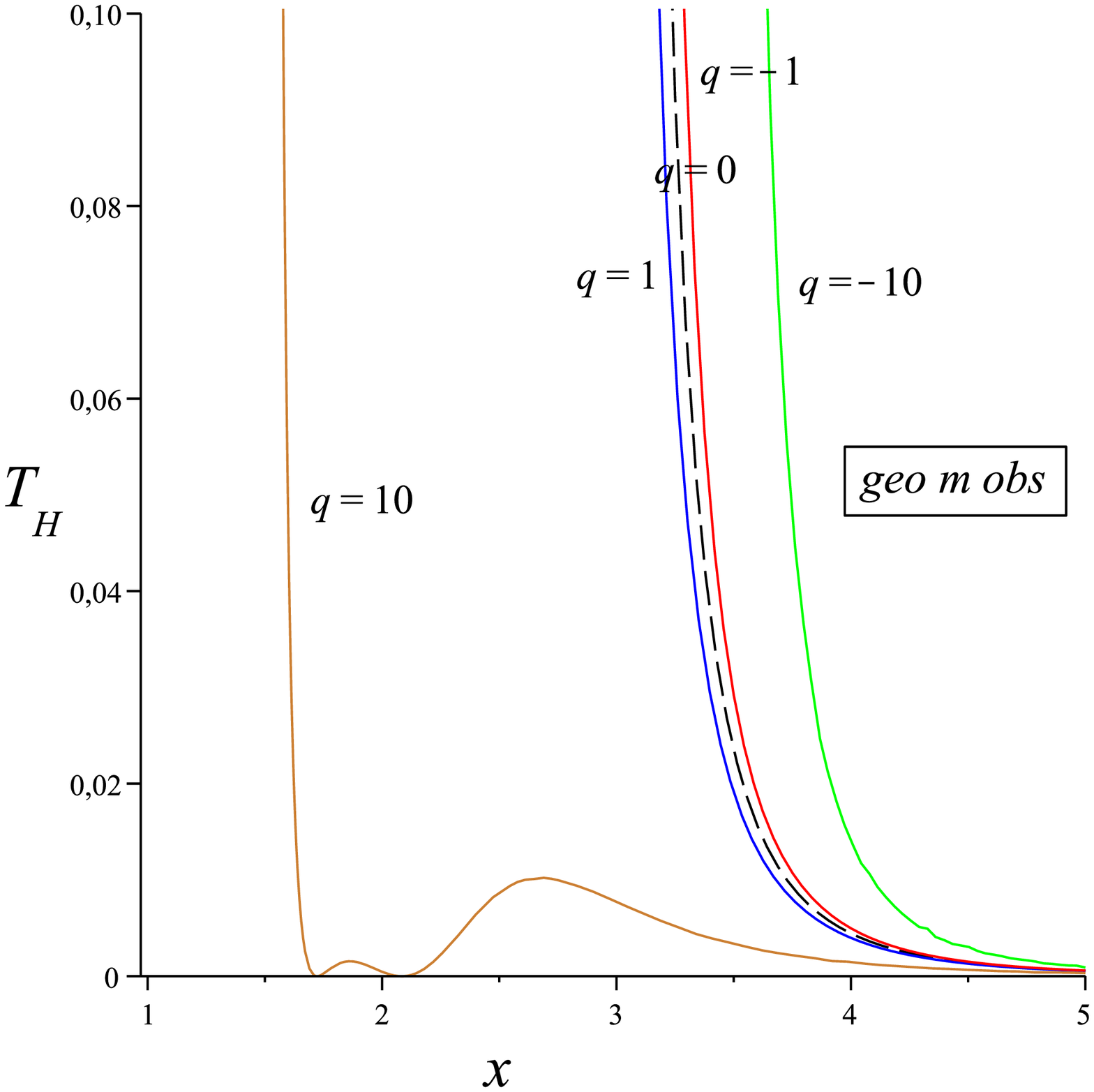}\\[.4cm]
\quad\mbox{(c)}\quad &\quad \mbox{(d)}\\
\end{array}$
\end{center}
\caption{The behaviors of the tidal indicators ${\mathcal T}_E(U)$ and ${\mathcal T}_H(U)$ as measured by geodesic observers are shown as functions of the distance parameter for the choice $a/M=0.5$ and different values of $q=[-10,-1,0,1,10]$.
Panels (a)--(b) and (c)--(d) correspond to the co-rotating and counter-rotating case respectively.
Dashed curves correspond to the Kerr case (i.e., $q=0$).
Units on the vertical axis are chosen so that $M=1$.   
}
\label{fig:7}
\end{figure}

\subsection{Limit of slow rotation and small deformation}

Let us consider the limiting case of the Hartle-Thorne metric written in terms of the more familiar Boyer-Lindquist coordinates according to the transformation (\ref{trasftoBL}).
Terms of the order $a^3$, $q^2$, $aq$ and higher are then neglected in all formulas listed below.
The relevant nonvanishing ZAMO frame components of the electric and magnetic parts of the Riemann tensor are given by
\begin{eqnarray}\fl\quad
E(n)_{11}&=&-\frac{2M}{r^3}-\frac{3a^2M}{r^5}N^2
+q\left[
\frac{6(r^2-3Mr+4M^2)}{Mr^3}Q_1
-\frac{3r-10M}{r^3}Q_2\right.\nonumber\\
\fl\quad
&&\left.+\frac{4M}{r^3}\ln{N}
\right]\,, \nonumber\\
\fl\quad
E(n)_{33}&=&\frac{M}{r^3}
+q\left[
\frac{3(7M^3-9rM^2+5r^2M-r^3)}{Mr^3(r-2M)}Q_1
+\frac{3r^2-11Mr+7M^2}{r^3(r-2M)}Q_2\right.\nonumber\\
\fl\quad
&&\left.-\frac{2M}{r^3}\ln{N}
\right]\,, \nonumber\\
\fl\quad
H(n)_{12}&=&-\frac{3aM}{r^4}N\,,
\end{eqnarray}
where $N=\sqrt{1-2M/r}$ and $Q_1=Q_1(r/M-1)$ and $Q_2=Q_2(r/M-1)$ are Legendre functions.
The Kretschmann invariant on the symmetry plane is
\beq\fl\quad
K=\frac{48M^2}{r^6}\left\{1-q\left[
\frac{6(r^2-3Mr+4M^2)}{M^2}Q_1
-\frac{3r-10M}{M}Q_2+4\ln{N}
\right]\right\}\,.
\eeq
Generalized Carter's observers have 4-velocity $\nu_*=aN/r$, whereas static observers are characterized by $\nu_{\rm(thd)}=-2aM/r^2N$ and geodesics observers by
\begin{eqnarray}\fl\quad
\nu_{\rm(geo)\pm}&=&\pm\sqrt{\frac{M}{r}}\frac{1}{N}\left[1+\frac{a^2}{2N^2r^4}(r^2+2Mr-12M^2)\right]-\frac{3aM}{r^2N}\nonumber\\
\fl\quad
&&\mp\frac32q\frac{(r-M)^2}{N^3\sqrt{Mr^3}}\left[\left(\frac{r}{M}-1\right)Q_1-Q_2\right]\,.
\end{eqnarray}

We list below the expression for ${\mathcal T}_H(U)$ corresponding to the different families of observers considered above:
\begin{eqnarray}\fl\quad
{\mathcal T}_H(n)&=&\frac{18a^2M^2}{r^8}N^2\,,\nonumber\\
\fl\quad
{\mathcal T}_H(u_{\rm(thd)})&=&\frac{18a^2M^2}{r^8}\frac{1}{N^2}\,,\nonumber\\
\fl\quad
{\mathcal T}_H(u_*)&=&0\,,\nonumber\\
\fl\quad
{\mathcal T}_H(u_{\rm(geo)\pm})&=&\frac{18M^3}{r^5}\frac{N^2}{(r-3M)^2}\mp36aM^2\frac{N^2}{r^5}\sqrt{\frac{M}{r}}\frac{r-M}{(r-3M)^3}\nonumber\\
\fl\quad
&&+\frac{18a^2M^2}{r^7}\frac{1}{(r-3M)^4}(r^3+Mr^2-13M^2r+15M^3)\nonumber\\
\fl\quad
&&-q\left[\frac{72M^3}{r^5}\frac{N^2}{(r-3M)^2}\ln{N}\right.\nonumber\\
\fl\quad
&&\left.+\frac{18M}{r^6}\frac{1}{(r-3M)^3}(9r^4-60Mr^3+170M^2r^2-248M^3r+153M^4)Q_1
\right.\nonumber\\
\fl\quad
&&\left.-\frac{18M^2}{r^6}\frac{1}{(r-3M)^3}(5r^3-29Mr^2+73M^2r-69M^3)Q_2
\right]\,.
\end{eqnarray}
Note that ${\mathcal T}_H(n)$ and ${\mathcal T}_H(u_{\rm(thd)})$ do not depend on $q$, in this limit, in contrast to the general case.
This is a consequence of the series expansion in the parameters $a/M$ and $q$, so that terms of the order of $q(a/M)$ have also been neglected.

Finally, in the weak field limit $M/r\ll1$ the previous expressions have the following asymptotic forms (up to the order $(M/r)^{10}$)
\begin{eqnarray}\fl\quad
\label{THweak}
{\mathcal T}_H(n)&\simeq&\frac{18a^2M^2}{r^8}\left(1-\frac{2M}{r}\right)\,,\nonumber\\
\fl\quad
{\mathcal T}_H(u_{\rm(thd)})&\simeq&\frac{18a^2M^2}{r^8}\left(1+\frac{2M}{r}\right)\,,\nonumber\\
\fl\quad
{\mathcal T}_H(u_{\rm(geo)\pm})&\simeq&
\frac{18M^3}{r^7}\left\{1-3q+\frac{M}{r}\left(4+\frac{a^2}{M^2}-9q\right)+\frac{M^2}{r^2}\left(15+13\frac{a^2}{M^2}-\frac{214}{5}q\right)\right.\nonumber\\
\fl\quad
&&\left.\mp2\frac{a}{M}\sqrt{\frac{M}{r}}\left(1+\frac{6M}{r}+29\frac{M^2}{r^2}\right)\right\}\,.
\end{eqnarray}

\section{Multipole moments, tidal Love numbers and Post-Newtonian theory}

Although the merging of compact objects can be accurately modeled only by numerical simulations in full general relativity, there is a variety of analytical and semi-analytical approaches which allow to properly describe at least part of the coalescence process and to study the associated gravitational wave signals (see, e.g., Refs. \cite{fla-hin,hin1,damour1}).
The effect of the tidal interaction on the orbital motion and the gravitational wave signal is measured by a quantity
known as the tidal Love number of each companion.

In Newtonian gravity, where it has been introduced \cite{love}, the Love number is a constant of proportionality between the external tidal field applied to the body and the resulting multipole moment of its mass distribution.
In a relativistic context instead Flanagan and Hinderer \cite{fla-hin,hin1} estimated the tidal responses of a neutron star to the external tidal solicitation of its companion, showing that the Love number is potentially measurable in gravitational wave signals from the early regime of the inspiral through Earth-based detectors.

The relativistic theory of Love numbers has been developed by Binnington and Poisson \cite{Bin-poi} and Damour and Nagar \cite{damour2}.
They classified tidal Love numbers into two types: an electric-type Love number having a direct analogy with the Newtonian one, and a magnetic-type Love number with no analogue in Newtonian gravity, already introduced in Post-Newtonian theory by Damour, Soffel and Xu \cite{dsxu}.
Damour and Lecian \cite{dlecian} also defined a third class of Love numbers, i.e., the ``shape'' Love numbers, measuring the distortion of the shape of the surface of a star by an external gravito-electric tidal field.
The relativistic Love numbers are defined within the context of linear perturbation theory, in which an initially
spherical body is perturbed slightly by an applied tidal field.
Tidal fields are assumed to change slowly with time, so that only stationary perturbations are considered. 
Computing the Love numbers requires the construction of the metric also in the interior of the body and its matching with the external metric at the perturbed boundary of the matter distribution.
Therefore, the internal problem depends on the choice of the stellar model, i.e., on the selected equation of state, whereas the external problem applies to a body of any kind.

Consider a massive body placed in a static, external tidal gravitational field, which is characterized by the
electric part of the associated Riemann tensor.
This tidal field will deform the body which will develop in response a gravitational mass quadrupole moment (and higher moments).
The components of the tidal field, quadrupole moment and total mass of the body will enter as coefficients the power series expansion of the spacetime metric in the body local asymptotic rest frame \cite{thorne,hin1}.
For an isolated body in a static situation these moments are uniquely defined. 
They are just the coordinate-independent moments defined by Geroch and Hansen \cite{ger,hans} for stationary, asymptotically flat spacetimes (see Eqs. (\ref{elemm})--(\ref{magmm})).

Tidal effects in relativistic binary system dynamics have been recently investigated in the framework of Post-Newtonian theory (see Ref. \cite{bfd} and references therein).
They have computed tidal indicators of both electric and magnetic types within the so called ``effective one body approach'' suitably modified to include tidal effects in the formalism, so improving the analytical description of the late inspiral dynamics with respect to previous works (see, e.g., Ref. \cite{damour1}).
In order to relate our results with the analysis done in Ref. \cite{bfd}, consider, for instance, the electric-type tidal indicator.
In the case of geodesic orbits and in absence of rotation we find in the weak field limit the following approximate expression (up to terms of the order $(M/r)^8$) 
\beq
{\mathcal T}_E(u_{\rm(geo)\pm})\simeq \frac{6M^2}{r^6}\left[1-2q+\frac{3M}{r}(1-q)\right]
\,,
\eeq
which can then be rewritten passing to harmonic coordinates $r=r_h+M$ and restoring the physical mass parameter \cite{ht67}, i.e., ${\mathcal M}=M(1-q)$, as
\beq
{\mathcal T}_E(u_{\rm(geo)\pm})\simeq \frac{6{\mathcal M}^2}{r_h^6}\left[1-\frac{3{\mathcal M}}{r_h}\right]\,.
\eeq
The same quantity (termed $J_a$ in Ref. \cite{bfd}, see Eq. (4.14) there) as above for the binary system at 1 PN order reads as 
\beq
{\mathcal T}_E(u_{\rm(geo)\pm})\simeq\frac{6{\mathcal M}^2X_2^2}{r_h^6}\left[1+\frac{(X_1-3){\mathcal M}}{r_h}\right]\,,
\eeq
where evaluation is performed in the center of mass system using harmonic coordinates and the mass of the two bodies are encoded in the parameters $X_1=m_1/{\mathcal M}$ and $X_2=m_2/{\mathcal M}$, with ${\mathcal M}=m_1+m_2$. 
In the limit $X_1=0$ ($X_2=1$) the two expressions coincide, as expected.

\section{Concluding remarks}

We have discussed the observer-dependent character of tidal effects associated with the electric and magnetic parts of the Riemann tensor with respect to an arbitrary family of observers in a generic spacetime.
Our considerations have then been specialized to the Quevedo-Mashhoon solution describing the gravitational field of a rotating deformed mass and to the family of stationary circularly rotating observers on the equatorial plane.
This family includes static, ZAMOs and geodesic observers and for each of them we have evaluated certain tidal indicators built up through the electric and magnetic parts of the Riemann tensor.
The main difference from the Kerr case examined in a previous paper is due to the presence of a genuine quadrupolar structure of the background solution adopted here: the total quadrupole moment of the source is not depending on the rotation parameter only, but there is also a further contribution due to the shape deformation directly related to the mass through a new mass quadrupole parameter, $q$.  
The properties of tidal indicators strongly depend on this new parameter.
We have found that there exists a family of circularly rotating orbits associated with $\nu=\nu_*$ along which the magnetic tidal indicator vanishes identically as in the Kerr case, playing the same role as Carter's observers there.
For special values of $q$ this property is also shared by other observer families, a novelty in comparison with the Kerr case.
However, still no observer family can be found for which the electric tidal indicator vanishes, a fact that can be explained in terms of curvature invariants.
The tidal electric indicator can be but extremized several times close to the source, showing also a damped oscillating behavior.    

We have also investigated the relation between tidal indicators and Bel-Robinson tensor, i.e., observer-dependent super-energy density and super-Poynting vector.
We have shown that the super-Poynting vector identically vanishes for $\nu=\nu_*$ leading to minimal gravitational super-energy as seen by such a generalized Carter's observer within the family of all circularly rotating observers at each spacetime point, a property already known to characterize Carter's observers in the case of black hole spacetimes.

\section*{Acknowledgements}
The authors thank Profs. R. Ruffini and T. Damour for useful discussions and ICRANet for support.

\appendix

\section{Relevant frame components of tidal tensors}

We list below the relevant nonvanishing ZAMO frame components of the electric and magnetic parts of the Riemann tensor:
\begin{eqnarray}\fl
E(n)_{11}&=&\frac{e^{-2\gamma}}{4T\sigma^2x^2}\left\{
\frac{X^2Sf_x^3}{2xf^2}
+\frac{X}{x}f_x^2f\omega(X\omega_x+x\omega)
-2X(Sf_{xx}+2f^3\omega\omega_{xx})\right.\nonumber\\
\fl
&&\quad\left.
-f_x\left[10Xf^2\omega\omega_x+\frac{2}{X}f^2\omega^2(X+2)+\frac{X}{2\sigma^2x}(4\sigma^4+Sf^2\omega_x^2)\right]\right.\nonumber\\
\fl
&&\quad\left.
-\frac{X}{x\sigma^2}f^3\omega\omega_x(f^2\omega_x^2-4\sigma^2)
-\frac{f^3}{\sigma^2}\omega_x^2(S+f^2\omega^2)
\right\}\,, \nonumber\\
\fl
E(n)_{33}&=&\frac{e^{-2\gamma}}{4\sigma^4fx^2}\left[\sigma^2f_x(2xf-Xf_x)+\omega_x^2f^4
\right]\,, \nonumber\\
\fl
H(n)_{12}&=&\frac{e^{-2\gamma}\sqrt{X}}{8Tx^3\sigma^5f}\left\{
2\omega\sigma^4[Xf_x^2(f_xX+xf)-6f^2f_x-2Xf^2(f_x+2xf_{xx})]
-f^5\omega_x^3S\right.\nonumber\\
\fl
&&\quad\left.
+f\sigma^2[-2f^3\omega\omega_x^2(Xf_x+3xf)+4Sf^2(\omega_x-x\omega_{xx})\right.\nonumber\\
\fl
&&\quad\left.+Sf_x\omega_x(Xf_x-10xf)]
\right\}\,,
\end{eqnarray}
where $X=x^2-1$, $S=\sigma^2X+f^2\omega^2$ and $T=S-2\sigma^2X$.

Furthermore, the Kretschmann invariant (\ref{Kdef}) of the QM spacetime evaluated on the equatorial plane $y=0$ is given by 
\begin{eqnarray}\fl
K&=&e^{-4\gamma}\frac{4X^2}{\sigma^4x^4}\left\{
f_{xx}^2-\frac{f^4\omega_{xx}^2}{X\sigma^2}
+\frac{f^2\omega_x\omega_{xx}}{2xX\sigma^4}[\sigma^2f_x(Xf_x-10xf)-f^2(f^2\omega_x^2-4\sigma^2)]
\right.\nonumber\\
\fl
&&\quad\left.
+f_{xx}\left[
(X+3)\frac{f_x}{Xx}-\frac{f_x^2}{2f^2x}(Xf_x+xf)+\frac{f^2\omega_x^2}{2Xx\sigma^2}(Xf_x+3xf)
\right]
\right.\nonumber\\
\fl
&&\quad\left.
+\frac{f_x^3}{16\sigma^2x^2f^2}\left[\frac{X\sigma^2}{f^2}f_x^2(Xf_x+2xf)-3f^2\omega_x^2(Xf_x-4xf)-8\sigma^2f_x\right.\right.\nonumber\\
\fl
&&\quad\left.\left.
-\frac{8x\sigma^2f}{X}(2X+3)\right]
+\frac{f^3f_x\omega_x^2}{2xX\sigma^2}\left(-\frac{7f^2\omega_x^2}{4\sigma^2}+13+\frac{6}{X}\right)
\right.\nonumber\\
\fl
&&\quad\left.
+f_x^2\left[
\frac{3f^4\omega_x^4}{16\sigma^4x^2}-\frac{f^2\omega_x^2}{4X\sigma^2x^2}(29X+25)+\frac{X^2+3X+3}{X^2x^2}
\right]
\right.\nonumber\\
\fl
&&\quad\left.
-\frac{f^4\omega_x^2}{X\sigma^2x^2}\left[
\frac{f^4\omega_x^4}{16\sigma^4}+1-\frac{f^2\omega_x^2}{4X\sigma^2}(5X+3)
\right]
\right\}\,.
\end{eqnarray}

\end{document}